\documentclass[prd,twocolumn,amsmath,amssymb,floatfix,superscriptaddress,nofootinbib,preprintnumbers]
{revtex4-1}
 \usepackage[utf8]{inputenc}
\usepackage{graphicx}
\usepackage{amssymb}
\usepackage{amsmath}
\usepackage{bm}
\usepackage{color}
\usepackage{enumitem}
\usepackage{booktabs}

\newcommand{\matr}[1]{\mathbf{#1}} 
\newcommand{\ra}[1]{\renewcommand{\arraystretch}{#1}}

\DeclareMathAlphabet\mathbfcal{OMS}{cmsy}{b}{n}
\definecolor{darkgreen}{RGB}{50,150,0}
\definecolor{purple}{cmyk}{0.5,0.75,0,0}
\definecolor{darkpurple}{RGB}{128,0,128}
\definecolor{ultramarine}{rgb}{0.07, 0.04, 0.56}
\definecolor{cadmiumgreen}{rgb}{0.0, 0.42, 0.24}
\definecolor{indigo(dye)}{rgb}{0.0, 0.25, 0.42}

\usepackage[linktocpage=true]{hyperref}
\hypersetup{
colorlinks=true,
citecolor=ultramarine,
linkcolor=cadmiumgreen,
urlcolor=indigo(dye),
pdfauthor={},
pdftitle={},
pdfsubject={}
}

\def\apj{{\rm ApJ}}                 
\def\mnras{{\rm MNRAS}}             
\def\prd{{\rm Phys.~Rev.~D}}        

\def\mnras{Mon.\ Not.\ R.\ Astron.\ Soc.\ }

\begin{document}
\preprint{YITP-SB-16-18}

\title{Model-Independent Predictions for Smooth Cosmic Acceleration Scenarios}

\author{Vinicius Miranda}

\affiliation{Center for Particle Cosmology, Department of Physics and Astronomy,\\University of Pennsylvania, Philadelphia, Pennsylvania 19104, USA}

\author{Cora Dvorkin}

\affiliation{Harvard University, Department of Physics, \\Cambridge, MA 02138, USA}

\begin{abstract}
Through likelihood analyses of both current and future data that constrain both the expansion history of the universe and the clustering of matter fluctuations, we provide falsifiable predictions for three broad classes of models that explain the accelerated expansions of the universe: $\Lambda$CDM, the quintessence scenario and a more general class of smooth dark energy models that can cross the phantom barrier $w(z)=-1$. Our predictions are model independent in the sense that we do not rely on a specific parametrization, but we instead use a principal component (PC) basis function constructed a priori from a noise model of supernovae and Cosmic Microwave Background observations. For the supernovae measurements, we consider two type of surveys: the current JLA and the upcoming WFIRST surveys. We show that WFIRST will be able to improve growth predictions in curved models significantly. The remaining degeneracy  between spatial curvature and $w(z)$ could be overcome with improved measurements of $\sigma_8 \Omega_m^{1/2}$, a combination that controls the amplitude of the growth of structure. We also point out that a PC-based Figure of Merit reveals that the usual two-parameter description of $w(z)$ does not exhaust the information that can be extracted from current data (JLA) or future data (WFIRST).
\end{abstract}

\maketitle

\section{Introduction}

The source of the current accelerated expansion of the universe, discovered almost two decades ago \cite{Riess:1998cb, Perlmutter:1998np}, remains one of the most intriguing puzzles of our time.
From an exotic component with negative pressure to a break of General Relativity on cosmological scales, many explanations have been theorized. 
Moreover, large  experimental efforts \cite{Riess:2006fw,Betoule:2014frx,2017arXiv170201747H,Eisenstein:2005su,Abell:2009aa,Amendola:2016saw,Ade:2015rim,Ade:2015zua,Benitez:2014ibt,2013MNRAS.433.2545E,2013ExA....35...25D,Riess:2017lxs,Abbott:2005bi,2017arXiv170405858A} either have been made, are in progress, or are currently  being proposed to measure the expansion history of the universe and growth of structure with percent-level precision (or better). 

Various parametrizations of the dark energy equation of state have been very well studied in the literature \cite{Huterer:2004ch,Wang:2005yaa,Zunckel:2007jm,Sullivan:2007pd,Zhao:2007ew,Zhao:2009ti,2009PhRvD..80l1302S,Kujat:2001ke,Chongchitnan:2007eb,Huterer:2006mv,Zhan:2008jh,Aghamousa:2016zmz}. In this work, we study a broader class of cosmic acceleration scenarios, modeling the equation of state $w(z)$ by a principal component (PC) basis function, following previous works \cite{Mortonson:2008qy,2010PhRvD..81f3007M,Vanderveld:2012ec}. We analyze a broad class of scenarios with a constant or time-dependent (but smooth) equation of state, with and without spatial curvature. This work provides both an update of the current state-of-the art on constraints on the Hubble expansion rate, as a function of redshift, the luminosity distance and the growth of structure, as well as predictions for the upcoming surveys. We use the constraints from current and future measurements to make predictions for other cosmic acceleration observables. Specifically, we use both current measurements of supernovae (JLA), as well as future measurements (WFIRST), observations of the Cosmic Microwave Background (CMB) temperature and polarization power spectra by the Planck satellite, baryon acoustic oscillations, and the Hubble constant. The results of this work can be used as a ground test for dark energy scenarios, a violation of which could potentially rule out a whole class of acceleration paradigms.

We will divide our cosmological observables in those providing us information about the geometry of the Universe, and those with information about the clustering of matter. In the context of smooth dark energy models, dark energy affects the growth of structures only through the background expansion. This assumption enables a consistency check by comparing observables that are sensitive to the background expansion and to the growth of linear perturbations, which is violated only in models that either modify general relativity or predict the clustering of the dark energy itself. Therefore, the rate of evolution of the growth functions with redshift is a powerful probe of dark energy~\cite{Cooray:2003hd,Ruiz:2014hma}. 
However, one should be cautious when interpreting consistency tests based on particular parametrizations because growth and geometry probes are sensitive to the evolution of the dark energy equation of state in different ways, and therefore wrong assumptions on the redshift behavior of $w(z)$ could induce misleading discrepancies. On the other hand, the choice of a particular parametrization has the advantage of being more computationally efficient, and can falsify interesting scenarios such as $w(z) = \text{constant}$. This paper offers a complementary approach to previous efforts~\cite{Cooray:2003hd,Ruiz:2014hma} that relied on specific functional forms for $w(z)$.

Another appealing possibility is to use $w(z)$ PCA to examine in detail how systematic effects and different survey strategies induce changes in the dark energy equation of state~\cite{2012PhRvD..86j3004R}. In the context of WFIRST, Ref. \cite{2017arXiv170201747H} provides a detailed analysis on the relative importance of various systematic uncertainties as well as the differences in the Figure of Merit between a space survey that carries an onboard integral field channel (IFC) spectrometer and a strategy that assumes that spectra will be observed from the ground. It would be interesting to understand how much their conclusions depend on the parametrization adopted, which we postpone for future work to be accomplished in collaboration with the WFIRST supernova science investigation teams. 

This paper is organized as follows. In \S\ref{sec:data} we discuss the data and broad classes of models that we use in our analysis. In \S\ref{sec:main_analysis} we present our main results, where we analyze the falsifiability of smooth dark energy scenarios ($\Lambda$CDM, quintessence, and more general smooth dark energy models that cross the phantom barrier of $w(z)=-1$), in a model-independent way.
In \S\ref{sec:FOM} we discuss the model-independent definition of Figure of Merit, first proposed in Ref. \cite{2010PhRvD..82f3004M}.
We then quantify, in \S\ref{sec:growth_probes}, the effects of marginalizing over spatial curvature on the different classes of dark energy scenarios studied, and viceversa (we see the effect on the curvature posteriors after marginalizing over the dark energy parameters). We present our conclusions in \S\ref{sec:discussion}.

\begin{figure}[t]
\vspace{1ex}
\includegraphics[scale=0.45]{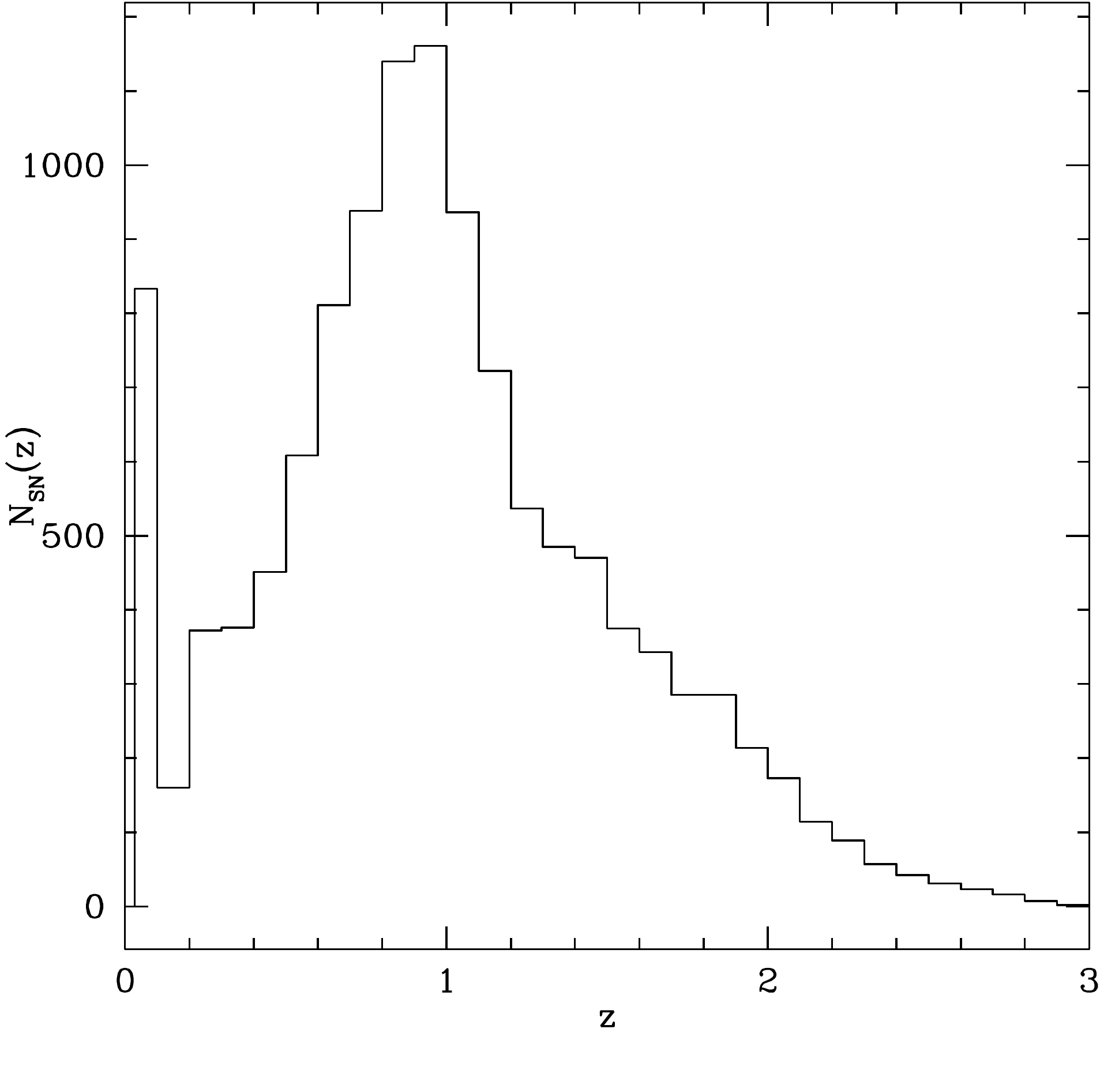}
\caption{Assumed type IA supernovae redshift distribution in the {\texttt Imaging-Allz} WFIRST strategy. The first bin ($0.01 < z < 0.1$) includes predictions for the number of type IA supernovae that will be observed by the Foundation supernovae survey \cite{2017AAS...22934112F,2017arXiv170201747H}.}
\label{fig:NZ_WFIRST}
\end{figure}

\section{Data and Models}
\label{sec:data}

\begin{figure*}[t]
\includegraphics[scale=0.45]{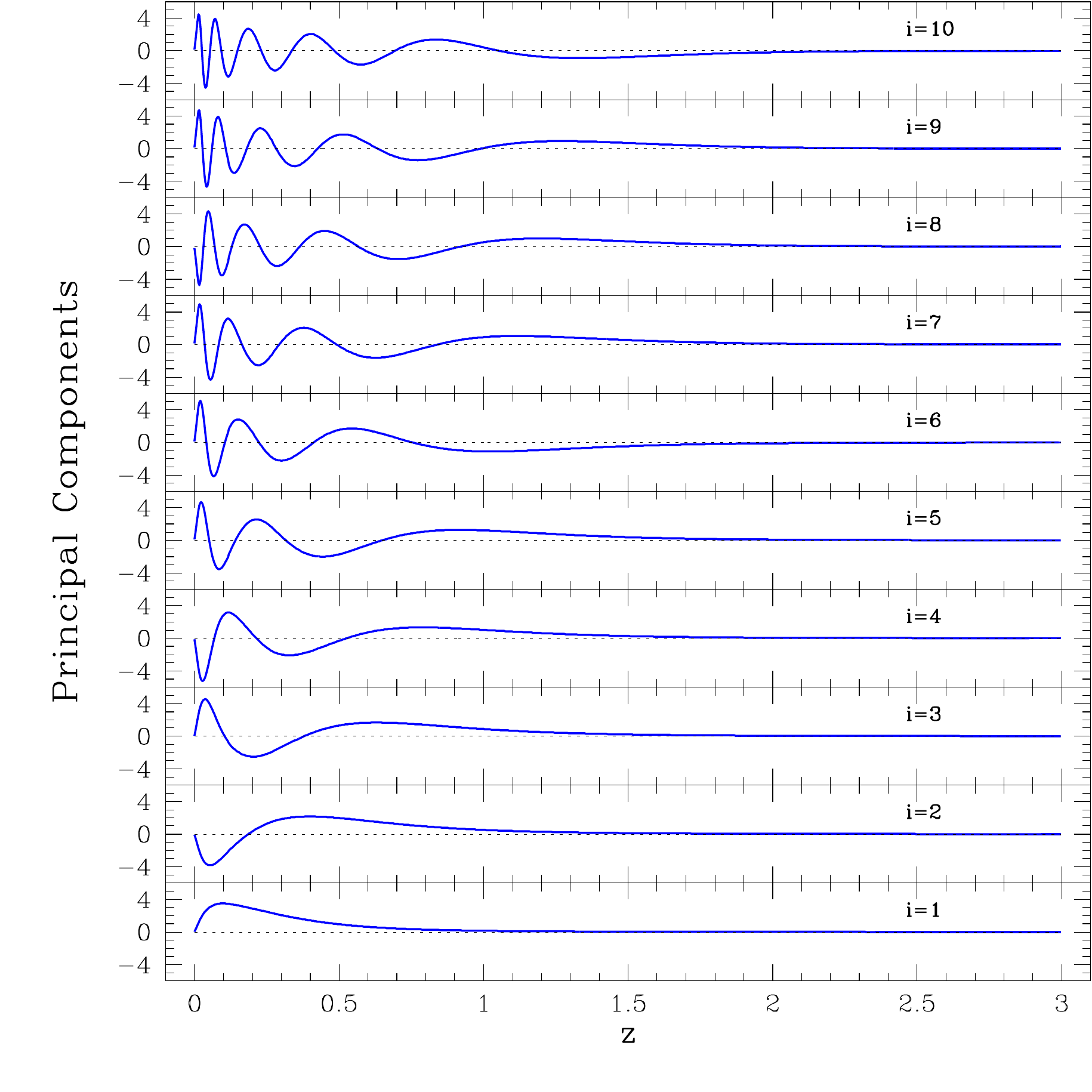}
\includegraphics[scale=0.45]{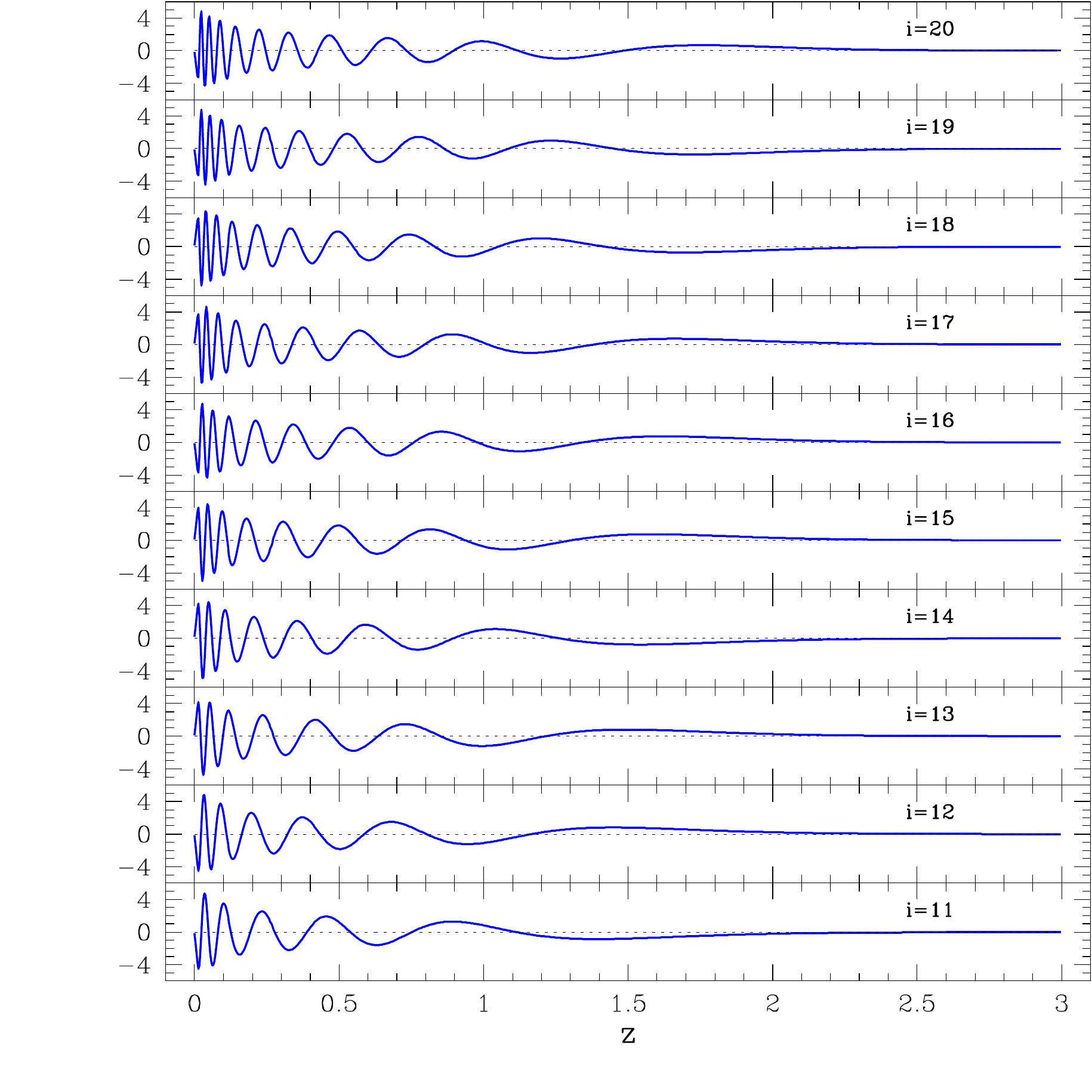}
\caption{The principal components of the dark energy equation of state, w(z). The Fisher matrix used to construct this basis contains contributions from both Type IA supernovae and the CMB. Lower components include fewer oscillations and have support, mainly, at low redshifts. Higher principal components oscillate rapidly at low redshifts, which suppresses their effect since there are two integrations in the scale factor to go from the equation of the state to the comoving luminosity distance. Their inclusion is, nevertheless, necessary to encompass the modes that supernovae can probe, with non-negligible statistical significance.}
\label{fig:PCA_WFIRST}
\end{figure*}

We ran multiple Markov Chain Monte Carlo (MCMC) likelihood analysis with a modified version of the CosmoMC code \cite{Lewis:2013hha,Lewis:2002ah,Lewis:1999bs,Howlett:2012mh}. Our chains were divided into three broad categories (see Tables~\ref{table:datasets_geo},~\ref{table:datasets_all} and~\ref{table:datasets_reduced}): the first group, named  {\it Geo}, contains datasets that probe only the geometry of our universe.  The second group, called {\it All}, also includes datasets that measure the linear and the nonlinear evolution of the structure formation. The last group, named {\it Reduced}, is similar to {\it Geo} but we exchange the CMB compressed Gaussian likelihood \cite{Ade:2015rim}\footnote{In the {\it Geo} chains, we have adopted the Gaussian compressed likelihood (not marginalized over the $A_L$ parameter).} with the full CMB temperature and polarization power spectra measurements \cite{Aghanim:2015wva}. 

\begin{table}[h]\centering
\ra{1.3}
\begin{tabular}{@{}rrrrcrrrcrrr@{}}\toprule
CMB &  \multicolumn{3}{c}{Gaussian} \\
BAO & \multicolumn{3}{c}{DR12+WiggleZ+6DF+MGS}\\
$H_0$ & \multicolumn{3}{c}{Riess et al. 2016} \\
SN & \multicolumn{3}{c}{JLA or WFIRST} 
\\
WL  & \multicolumn{3}{c}{ -- } \\ \bottomrule
\end{tabular}
\caption{Datasets that define the {\it Geo} group of chains.}
\label{table:datasets_geo}
\end{table}

In this paper, we have adopted the {\it Reduced} dataset mainly to assess how the discrepancies between low-redshift probes and the CMB affects the measurement of curvature in the different dark energy scenarios.

\begin{table}[h]\centering
\ra{1.3}
\begin{tabular}{@{}rrrrcrrrcrrr@{}}\toprule
CMB & \multicolumn{3}{c}{Full Planck (includ.  lensing reconstruction)} \\
BAO & \multicolumn{3}{c}{DR12 (includ. RSD)+WiggleZ+6DF+MGS}\\
$H_0$ & \multicolumn{3}{c}{Riess et al. 2016} \\
SN &  \multicolumn{3}{c}{JLA} \\
WL & \multicolumn{3}{c}{CFHTLens (includ. non-linear scales)} \\ \bottomrule
\end{tabular}
\caption{Datasets that define the {\it All} group of chains.}
\label{table:datasets_all}
\end{table}

In addition to these broad categories, our chains in the {\it Geo} group were further divided into two sub-groups, depending on the adopted supernovae dataset. The first sub-group uses the current JLA compilation implemented on CosmoMC \cite{Betoule:2014frx}, and the second one adopts WFIRST simulated data \cite{2017arXiv170201747H}. 

All the chains ran in this work included BAO \cite{2016MNRAS.460.4210G, Blake:2011en} and local $H_0$ measurements \cite{Riess:2016jrr}. The local $H_0$ measurements are implemented in CosmoMC as a Gaussian prior in the inverse angular diameter distance at the effective redshift $z=0.04$~\cite{2009ApJ...699..539R}; therefore there is a dependency on the dark energy equation of state, which is taken into account. Indeed, we see some broadening in the $\chi^2$ posterior as a function of $H_0$ in the smooth dark energy models when compared to $\Lambda$CDM.

\begin{table}[h]\centering
\ra{1.3}
\begin{tabular}{@{}rrrrcrrrcrrr@{}}\toprule
CMB &   \multicolumn{3}{c}{Full Planck}\\
BAO &  \multicolumn{3}{c}{DR12+WiggleZ+6DF+MGS}\\
$H_0$ & \multicolumn{3}{c}{Riess et al. 2016} \\
SN &  \multicolumn{3}{c}{JLA} \\
WL&  \multicolumn{3}{c}{ -- } \\ \bottomrule
\end{tabular}
\caption{Datasets that define the {\it Reduced} group of chains}
\label{table:datasets_reduced}
\end{table}

Among all the different strategies presented in the WFIRST supernovae analysis \cite{2017arXiv170201747H}, we adopted the so-called {\texttt Imaging-Allz}. That {\texttt Imaging-Allz} setup provides measurements of a few thousand type IA supernovae in a few redshift bins, observed over the broad redshift range $0 < z < 3$ (see Figure~\ref{fig:NZ_WFIRST}). It does so by considering the scenario that a ground-based spectroscopy will be sufficient to calibrate the redshift evolution of the supernovae spectral features, which allows the WFIRST satellite to be solely an imaging survey. Complications of this hypothesis work in the direction of lowering the number of observed supernovae, increasing the systematic errors, and shortening the redshift range. We intend to address the impact that the different WFIRST strategies have in our conclusions in future work.

The {\it All} set of chains constrain the evolution of perturbations by adding the full lensed Planck temperature and polarization data \cite{Aghanim:2015wva}, CMB lensing reconstruction \cite{Ade:2015zua}, redshift space distortions \cite{2016MNRAS.460.4210G}, and tomographic CFHTLenS weak lensing data \cite{Heymans:2013fya}. For both the CMB lensing reconstruction and the CFHTLenS datasets, we have used range cuts that are less conservative than the ones adopted by the Planck collaboration, as they include scales where gravitational collapse is non-linear \cite{Ade:2015rim}. On the lensing reconstruction likelihood, we adopted the so-called aggressive cuts, and on the CFHTLenS weak lensing, we adopted the six bin tomographic likelihood. With the potential systematic contamination in mind, we do not over-emphasize the statistical significance of deviations from $\Lambda$CDM.

\begin{figure}[t]
\vspace{1ex}
\includegraphics[scale=0.44]{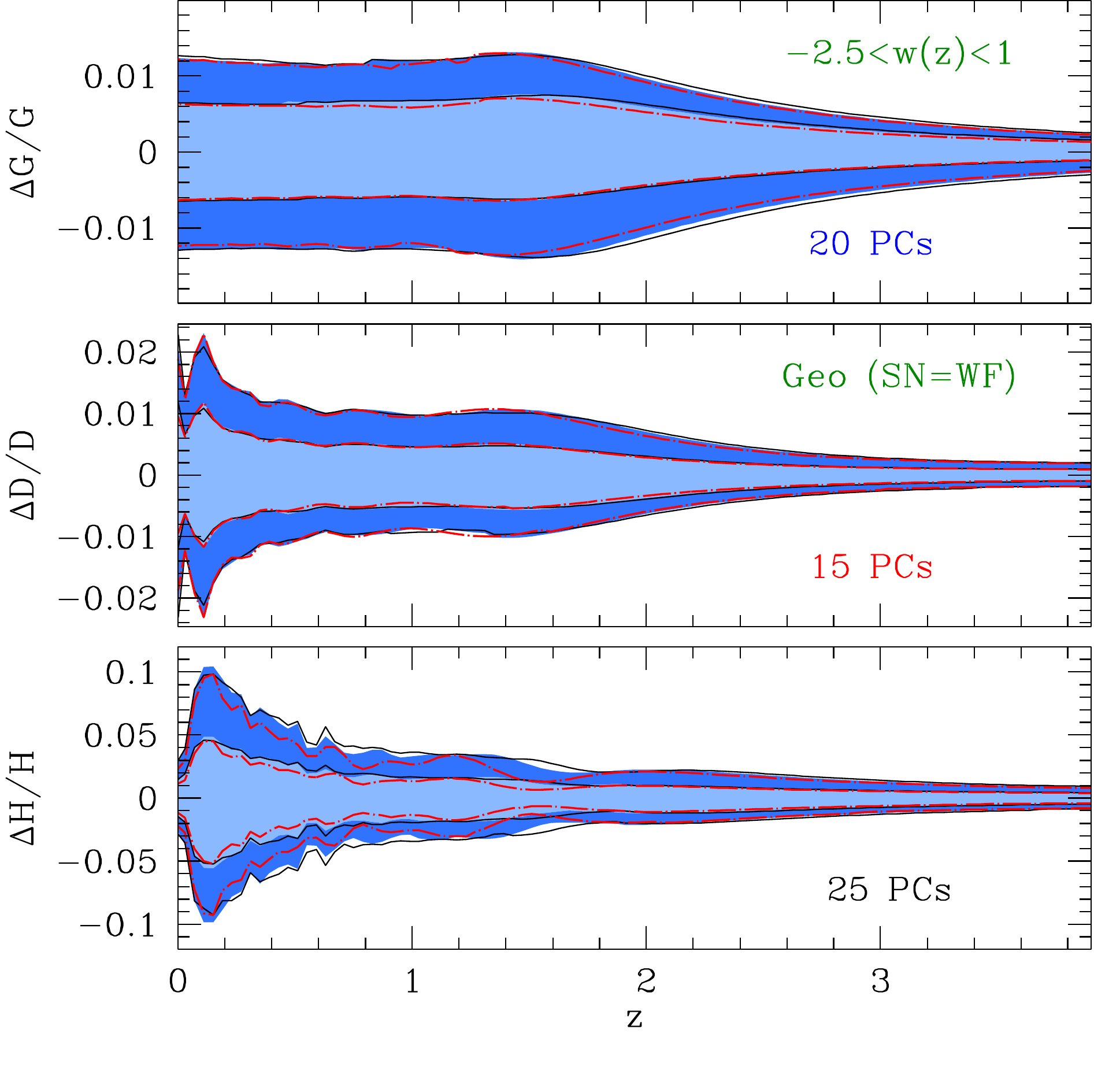}
\caption{Posterior for the grow function, the comoving luminosity distance and the Hubble expansion rate (top to bottom) predicted by chains with 15 (dot-dashed red lines), 20 (blue shaded areas) and 25 (solid black lines) principal components. The inner lines (and the darker blue shaded region) correspond to the width of the $68\%$ confidence interval, while the outer lines (and the lighter blue shades) indicate the width of the $95\%$ confidence area. All posteriors are centered at zero to guide the comparison of their width. In all three chains, we assumed flatness and the prior $-2.5<w(z)<1$.  The likelihoods adopted in these chains are shown in Table~\ref{table:datasets_geo}, with the supernovae data given by the simulated WFIRST data. The small differences in the Hubble posteriors are suppressed even further in the distance posteriors, which indicates that they reflect amplitude shifts in highly oscillatory modes. The small changes in distance and growth observables also confirm that $20$ PCs are sufficient to ensure completeness of the PCA basis.}
\label{fig:PCA_WFIRST_COMPLETE}
\end{figure}

To account for the non-linear scales in the matter power spectrum, we adopt the HALOFIT fit \cite{Takahashi:2012em}. Given that HALOFIT has only been calibrated to models with a constant dark energy equation of state, we apply the mapping described in Ref. \cite{Casarini:2016ysv} between a general time-evolving dark energy. equation of state and $w=\text{const}$.  This mapping has been tested against simulations for the well known $w_0-w_a$ dark energy parametrization, and we assume that the arguments presented in Refs. \cite{Casarini:2008np, Casarini:2016ysv} that justify this mapping are also valid here. Indeed, Ref.~\cite{Casarini:2008np} has argued that the non-linear completion (of the power spectrum) of rapidly varying $w(z)$ models can be obtained from $w=\text{const}$ models by matching the distance to the last scattering surface.

We fix the sum of neutrino mass as $\sum m_\nu = 0.06$eV in all the chains. In a future paper, we will analyze how the sum of neutrino masses is affected when marginalizing over different assumptions on the dark energy equation of state. 

\begin{figure*}[t]
\includegraphics[scale=0.44]{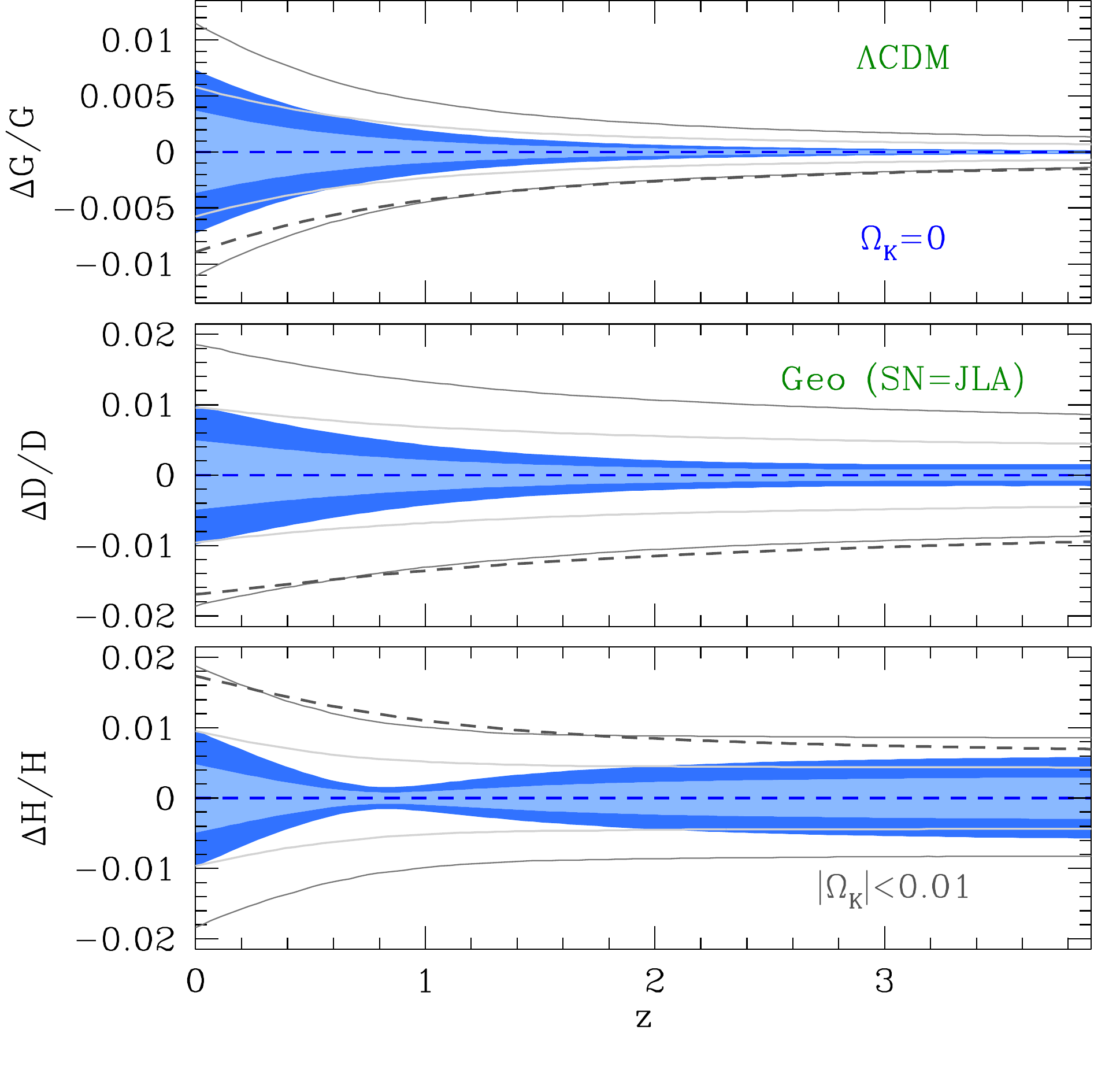}
\includegraphics[scale=0.44]{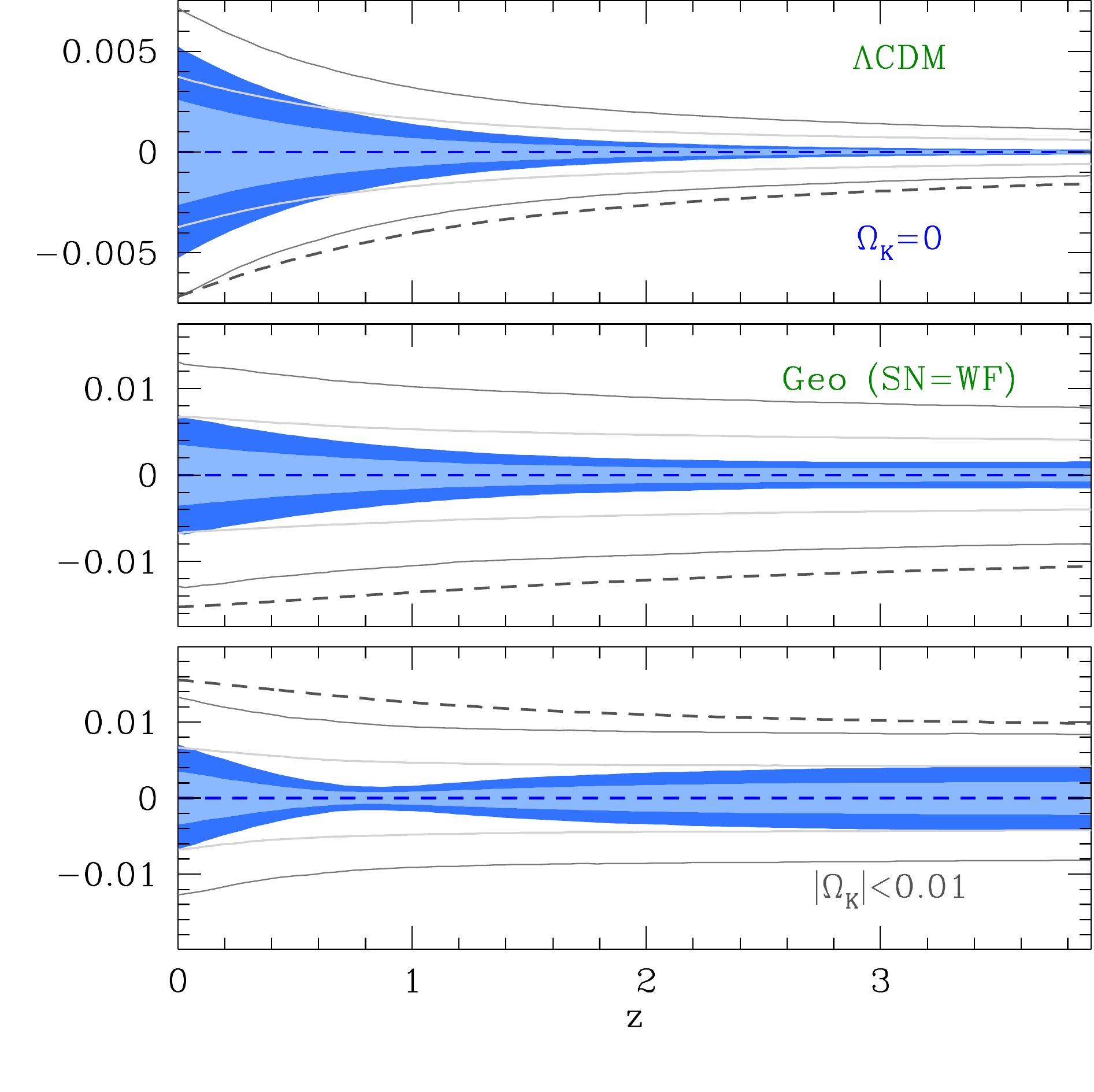}
\caption{Constraints on the growth function, the comoving luminosity distance, and the Hubble expansion rate, predicted in the $\Lambda$CDM scenario. The blue contours and the grey lines show the $68\%$ (light) and $95\%$ (dark) confidence region assuming no curvature ($\Omega_k = 0$) and the flat prior $ |\Omega_k| < 0.01$ (which is non-informative), respectively. In both cases, the confidence levels were centered at zero to highlight the broadening of the posterior. Dashed lines show the fractional difference of the posterior means to flat $\Lambda$CDM. The likelihoods adopted in these chains are shown in Table~\ref{table:datasets_geo}, with the supernovae given by the JLA compilation on the left panel and the WFIRST simulated data on the right panel. For flat $\Lambda$CDM, the remaining freedom on the growth function and the comoving luminosity distance is less than a percent on the entire redshift range, even though the Hubble constant is measured at the $2.4\%$ level. As noted by Ref. \cite{Mortonson:2008qy}, flat $\Lambda$CDM predictions on the Hubble expansion rate are especially tight around $z\approx 1$, which opens an interesting window of opportunity for model testing by future experiments.}
\label{fig:LCDM1} 
\end{figure*}

In this paper we give an up-to-date status of the falsifiability of smooth dark energy models beyond $\Lambda$CDM. Furthermore, we analyze the impact of marginalizing different dark energy scenarios on parameters like the curvature on spatial slicings, $\Omega_K$, and how measurements of a parameter that monitors local structure, $S_8 \equiv \Omega_m^{1/2} \sigma_8$, are relevant to constraining dark energy and $\Omega_K$ simultaneously\footnote{The parameter $\sigma_8$ is defined as the RMS amplitude in linear theory of mass fluctuations on $8 h^{-1}$ Mpc scale.}.  Instead of performing a case-by-case analysis, we chose to do a model-independent analysis using a principal component basis, being the main advantage of this basis the fact that it is complete.

We expand the dark energy equation of state as
\begin{align}
w(z) = w_\text{fiducial} + \sum_{i=1}^{N_\text{PC}} \alpha_i e_i(z),
\end{align}
where $e_i(z)$ with $i=1,...,N_\text{PC}$ are the principal components of perturbation around the fiducial model $w_\text{fiducial} = -1$. These are shown in Figure~\ref{fig:PCA_WFIRST}. The principal components have support in the range $0 < z < z_\text{max}=3$. For $z>z_\text{max}$, we extrapolate the equation of state by assuming $w = w_\infty = \text{constant}$. Therefore, the energy density of dark energy is given by
\begin{align}
\rho_\text{DE}(z) = 
\begin{cases}
\rho_\text{DE}(0)\exp\Big[3 \int_0^z dz' \frac{1+w(z')}{1+z'}\Big],  \quad z \leq z_\text{max} \\
\rho_\text{DE}(z_\text{max})  \Big(\frac{1+z}{1+z_\text{max}}\Big)^{3(1+w_\infty)},  \quad z > z_\text{max}.
\end{cases}
\end{align} 
Here, and throughout this paper, we assume $w_\infty = -1$.

The parameter vector in the {\it Geo} chains is 
$ \vec{\theta}_\text{Geo} = \{ \Omega_c h^2, \theta_A, \alpha_1,..., \alpha_{N_\text{PC}}, \Omega_K \}$. Here, $\Omega_c h^2$ is the cold dark matter density, $\theta_A$ is the angular size of the horizon at the time of recombination, $H_0$ is the local Hubble constant and $h \equiv H_0 \big/(100 \text{ km/s/Mpc})$. To reduce the dimensionality of the expensive MCMC chains,  we fixed the baryon density, $\Omega_b h^2 = 0.02228$ and the scalar tilt, $n_s = 0.966$ in all the {\it Geo} chains, even though there is a correlation between these two quantities with the angular size of the CMB peaks and the so-called shift parameter $R = \sqrt{\Omega_mH_0^2}D_A(z_*)/c$ \cite{Efstathiou:1998xx, Ade:2015rim}, where $D_A(z)$ is the comoving angular diameter distance to redshift $z$, and $z_*$ is the redshift of recombination. Most of the constraining power on the amplitude of the principal components comes, however, from type IA supernovae and not from the CMB.  

The baseline model for the chains in {\it All} and {\it Reduced} groups is
\begin{align}
\vec{\theta}_\text{All/Red} &= \vec{\theta}_\text{Geo}+\{\Omega_b h^2, n_s, \ln A_s, \tau\}. 
\end{align}
Here, $\tau$ is the reionization optical depth, $\Omega_b h^2$ is the baryon density, and $\log A_s$ and $n_s$ are the initial curvature power spectrum amplitude and tilt, respectively. The reionization history is assumed to be given by the so-called instantaneous reionization. Generalizations of the reionization history that better fit the Planck LFI polarization data could potentially affect our results by changing the inferred $\ln A_s$, which then affects the gravitational lensing amplitude \cite{Heinrich:2016ojb}.

The growth function obeys the following evolution equation,  
\begin{align}
G'' + \Bigg(4 + \frac{H'}{H} \Bigg) G' + \Bigg[ 3 + \frac{H'}{H} - \frac{3}{2} \Omega_m(z)  \Bigg] G = 0.
\end{align} 
Here $H(z)$ is the Hubble function (we have neglected radiation);  $\Omega_m(z) = \Omega_m (1+z)^2 [H_0/H(z)]^2$; and prime denotes derivative with respect to $\ln a$. The normalization of the growth function at the initial redshift $z_\text{ini}=1000$ is $G(z_\text{ini}) = 1$, and $G'(z_\text{ini}) = -6 \, \Omega_\text{DE}(z_\text{ini})/5$, where $\Omega_\text{DE}(z) = [\rho_\text{DE}(z)/\rho_\text{DE}(0)] [H_0/H(z)]^2$ \cite{Mortonson:2008qy}. 
Finally, the logarithm growth rate is defined as
\begin{align}
f(z) \equiv \frac{d\ln D}{d\ln a}= 1 + \frac{G'}{G}.
\end{align}
Often, the parametrization $f(z) = \Omega_m(z)^{\gamma}$ is assumed~\cite{Linder:2005in}, so we will also show how the so-called growth index $\gamma$ varies as function of the redshift.

To construct the PC basis, we closely follow the Appendix A of Ref. \cite{Mortonson:2008qy}. Here we will summarize the procedure and highlight the differences to Ref. \cite{Mortonson:2008qy}. We start with the supernovae Fisher matrix
\begin{align}
\matr{F}_{ij}^\text{SN} = \sum_\beta \sigma_\beta^{-2} \frac{d m(z_\beta)}{d p_i}\frac{d m(z_\beta)}{d p_j},
\end{align}
where $\beta$ runs through the redshift binning, $m$ is the apparent magnitude $m(z) \equiv \mathcal{M} + 5 \log(H_0 d_L(z))$, $\mathcal{M}$ is a constant related to the absolute magnitude, $d_L$ is the luminosity distance, $H_0$ is the Hubble constant, $p_i = \{\beta_1,...,\beta_{N_z}, \mathcal{M}, \Omega_m, \Omega_m h^2\}$, and $ \{\beta_1,...,\beta_{N_z} \}$ are the amplitudes of a binned dark energy equation of state.
The eigenvectors of the Fisher matrix generate a basis for arbitrary functions defined on the redshift bins, $w(z_j) = w_\text{fid} + \sum_{i=1}^{N_\text{z}} \alpha_i e_i(z_j)$. 
We construct piecewise-rectangular-shaped, $w(z) = \beta_i$ if $z_{i-1} < z < z_i$ (and zero otherwise), which reduces the numerical noise in the final PCA shape because the energy density can be evaluated analytically.

The model for the statistical and systematic errors adopted in the supernovae Fisher matrix was updated relative to \cite{Mortonson:2008qy} to better represent the WFIRST {\texttt Imaging-Allz} simulated data:
\begin{align}\label{eqn::SN_FISHER_ERROR}
\sigma_\beta^2 = \Big(\frac{\Delta z}{\Delta z_\text{sub}}\Big) \Bigg[ \frac{s(z)}{N_\beta} + 0.01^2  \Big(\frac{1+z}{1.7}\Big)^2 \Bigg],
\end{align}
with 
\begin{align}
s(z) = 
\begin{cases} 
      0.015 & z\leq 1.03 \\
      -0.014 + 0.014 \times (1+z) & z >1.03. \\
         \end{cases}
\end{align}
Here, $N_\beta$ is the number of supernovae in each bin and $\Delta z = 0.1$, except for the first bin where  $\Delta z = 0.1 - z_\text{min}$ with $z_\text{min}=0.01$. We subdivided the data into $N_{z} = 883$ sub-bins up to $z_\text{max}=3$ (making $\Delta z_\text{sub}=0.003$), which corresponds to the maximum observable supernovae redshift in the WFIRST {\texttt Imaging-Allz} strategy \cite{2017arXiv170201747H}. The number of supernovae in the bins of $\Delta z = 0.1$ is shown in Figure~\ref{fig:NZ_WFIRST}. As stated in Ref. \cite{2017arXiv170201747H}, the systematic model adopted in our Fisher matrix is an oversimplification. More realistic likelihoods will, in practice, introduce correlations between the principal components. This, however, does not affect the conclusions of this paper since they do not depend on the orthogonality of the basis. The crucial point is that the PCA basis, although not orthogonal, is complete, i.e, it contains all the modes that can be observed by the WFIRST likelihood with high statistical significance. Indeed, Figure~\ref{fig:PCA_WFIRST_COMPLETE} compares the posterior for the Hubble expansion rate, the growth function, and the comoving luminosity distance when $15$, $20$, and $25$ principal components are varied. From here, it is clear that going from $15$ to $25$ PCs does not alter the observable posterior significantly.

\begin{figure}[t]
\includegraphics[scale=0.44]{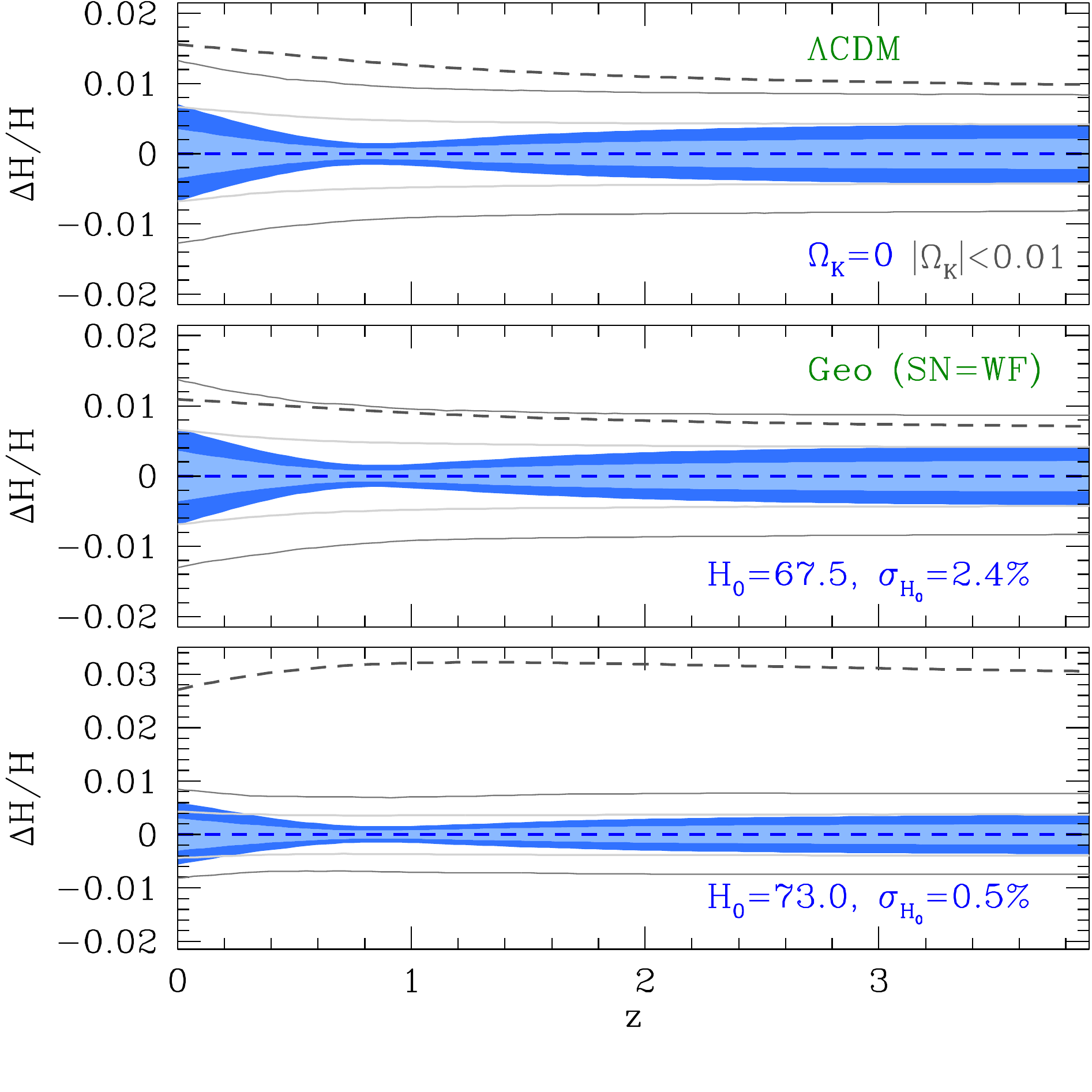}
\caption{Effects on the Hubble expansion rate predicted by $\Lambda$CDM, induced by the discrepancy between local $H_0$ measurements and the angular position of the CMB acoustic peaks. The blue contours and the solid grey lines show the $68\%$ (dark) and $95\%$ (light) confidence levels, assuming no curvature ($\Omega_k = 0$), and the prior $ |\Omega_k| < 0.01$, respectively. In both cases, the confidence levels were centered at zero to highlight the broadening of the posteriors. Dashed lines show the fractional difference of the posterior means relative to flat $\Lambda$CDM. Table~\ref{table:datasets_geo} explains the likelihoods adopted in these chains with the Type IA supernovae data from simulated WFIRST dataset. In the middle panel, we artificially shift the mean value of the local $H_0$ measurements value to $67.5$ km/s/Mpc. This displacement does not change the posterior width at a significant level, but does shift the mean prediction for the Hubble expansion rate in curved $\Lambda$CDM. Therefore, the width of our predictions is not artificially tighten as a result of the discrepancy between the CMB and local measurements.  Going to sub-percent $H_0$ measurements would make our analysis completely unreliable, as shown in the lower panel, in the case the source of discrepancy is systematic errors, or it could provide a significant detection of positive curvature, which could be falsified by measuring the amplitude of the growth of structure.}
\label{fig:test_H0} 
\end{figure}

\begin{figure*}[t]
\includegraphics[scale=0.44]{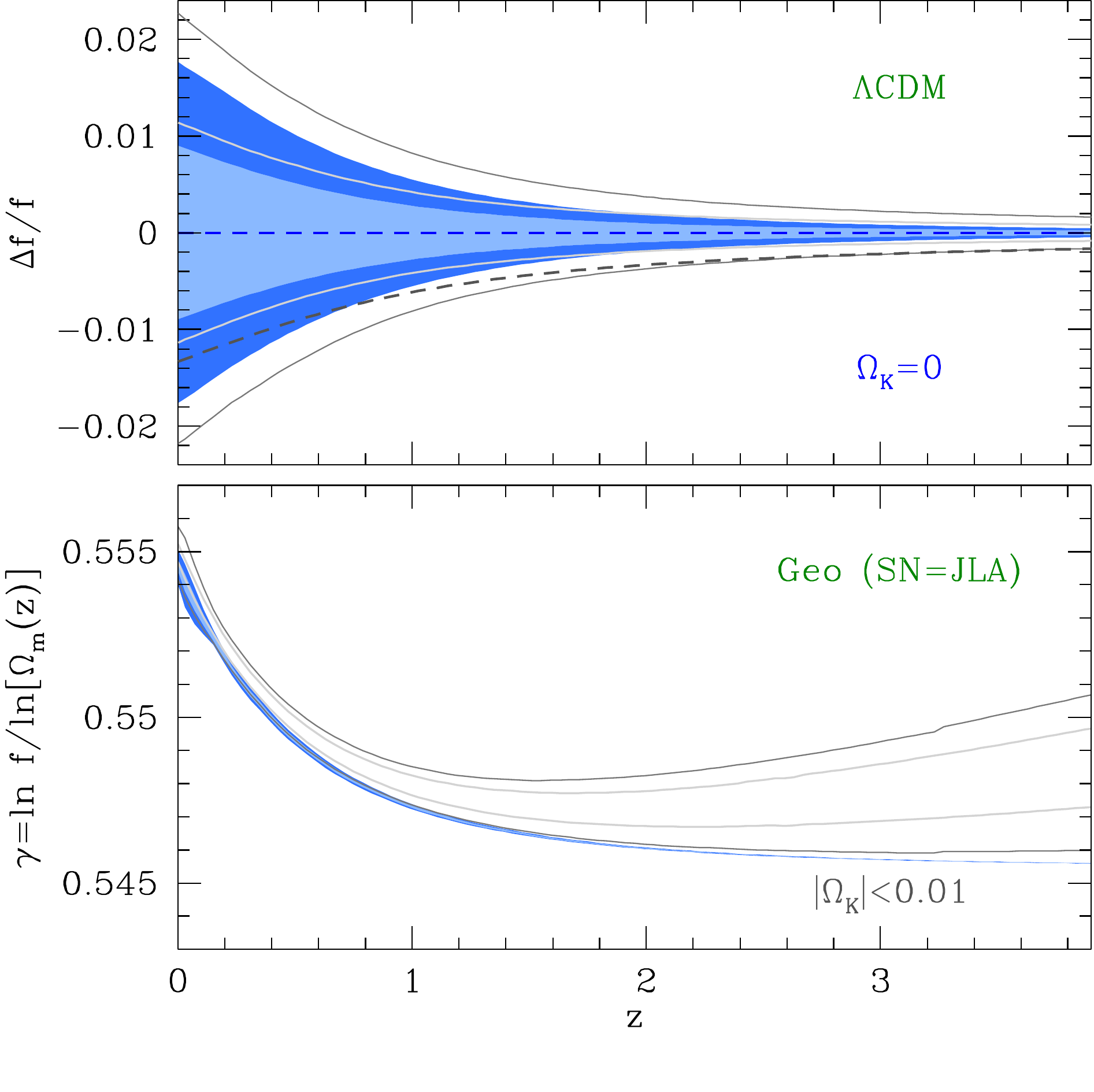}
\includegraphics[scale=0.44]{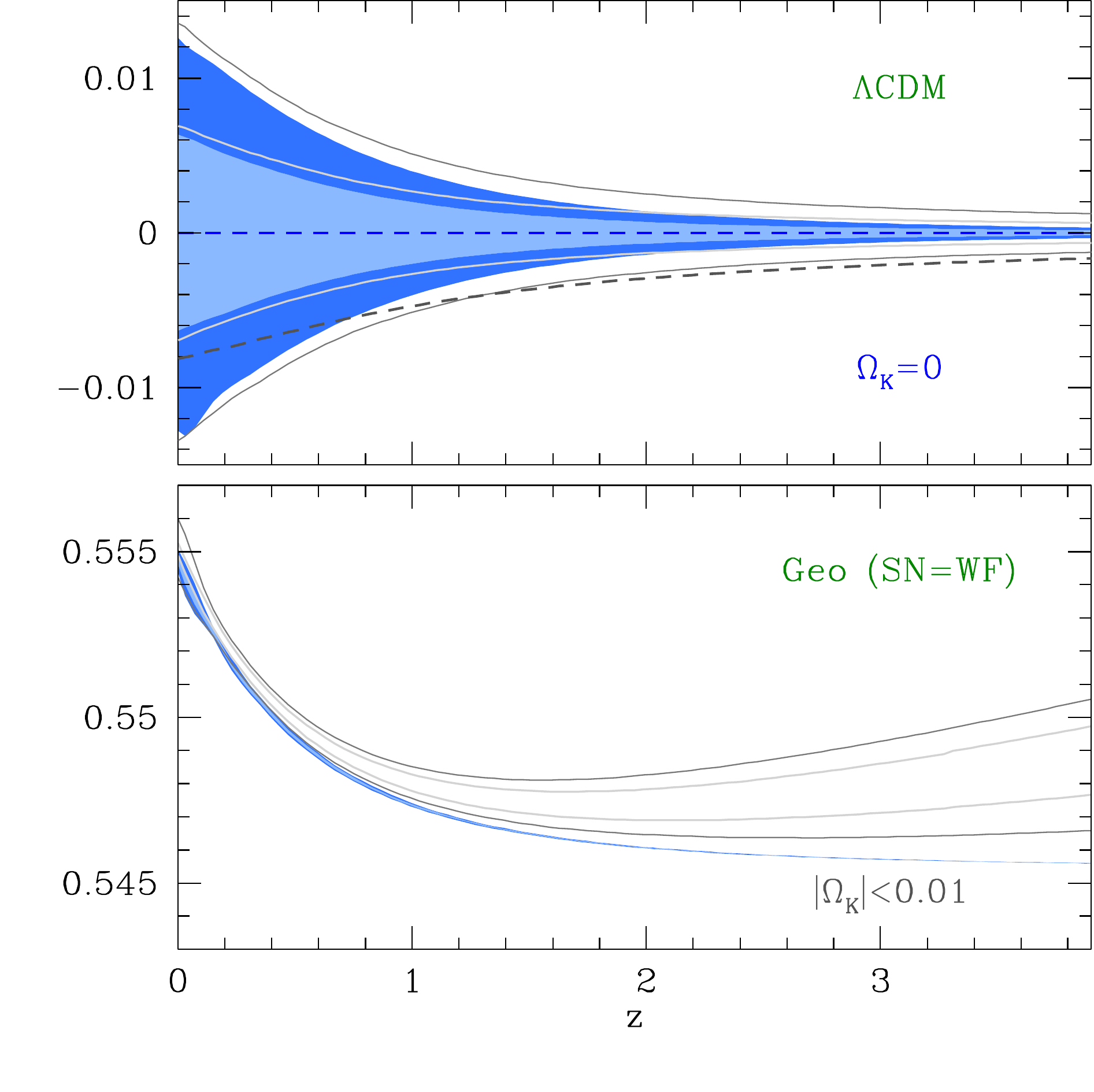}
\caption{Constraints on the growth rate and the growth index, predicted by $\Lambda$CDM models.  The blue contours show the $68\%$ (light) and $95\%$ (dark) confidence region assuming no curvature ($\Omega_k = 0$), while the solid grey lines assume the prior $ |\Omega_k| < 0.1$.   In both cases, the confidence levels were centered at zero to highlight the broadening of the posterior. Dashed lines show the fractional difference of the posterior means relative to flat $\Lambda$CDM. Table~\ref{table:datasets_geo} shows the likelihoods adopted in these chains, with the supernovae data given by the JLA compilation (left panel) and the WFIRST simulated data (right panel). Predictions for flat $\Lambda$CDM are so tight that the shades are barely visible in the lower panel.}
\label{fig:LCDM2} 
\end{figure*}

To smooth the shape of the principal components, we apply the continuum limit, i.e., we increase the corresponding number of equal size bins to $N_z$, and then we impose the normalization 
\begin{align}
\sum_{i=1}^{N_\text{z}} \, \, [e_i(z_j)]^2 = \sum_{j=1}^{N_\text{z}} \, \, [e_i(z_j)]^2 = N_z \, .
\end{align}
We calculate the number of supernovae in each sub-bin via linear interpolation from the center of each original redshift bin. The factor $\Delta z/\Delta z_\text{sub}$ rescales the errors in the sub-bins. Because the number of principal components ($N_\text{PC}$) that ensures completeness with the data is much less than total number of bins, $N_z$, none of our results will depend on sub-bin width $\Delta z_\text{sub}$. Last, we tested our procedure by explicitly reproducing the PCA basis shown in Ref. \cite{Mortonson:2008qy}.

We also add a Planck-like likelihood to the total Fisher matrix. Similar to Ref. \cite{Mortonson:2008qy}, we adopt the covariance matrix 
\begin{align}
\matr{C}^\text{CMB}=
\begin{bmatrix}
    (0.0018)^2   & -(0.0014)^2 \\
    -(0.0014)^2   & (0.0011)^2 \\
\end{bmatrix},
\end{align}
for the parameters $\vec{q} = \{\ln(D_\ast \big/\text{Mpc}), \Omega_m h^2\}$, where $D_\ast$ is the comoving distance to the surface of the last scattering\footnote{To evaluate $D_\ast$, we include radiation as well as cold dark matter and dark energy contributions.}. We then construct the CMB Fisher $\matr{F}^\text{CMB}= \matr{D} [\matr{C}^\text{CMB}]^{-1} \matr{D}^T$, where $\matr{D}_{ij} = d{q}_i \big/dp_j$. 

Therefore, the total Fisher matrix is $\matr{F} = \matr{F}^\text{SN} + \matr{F}^\text{CMB}$. We then marginalize $\matr{F}$ over $\mathcal{M}$, $\Omega_m$ and $\Omega_m h^2$. Supernovae measurements are, therefore, insensitive to constant shifts in relative distances as well as shifts that are nearly constant at $z>z_\text{min}$. As explained in Appendix B of Ref. \cite{Mortonson:2008qy}, large variations in $w(z)$ below $z<z_\text{min}$ create degeneracies between $\{\alpha_1,...,\alpha_{N_\text{PC}} \}$ and $\Omega_m$ that slow the convergence of the chains. 

\begin{figure*}[t]
\includegraphics[scale=0.44]{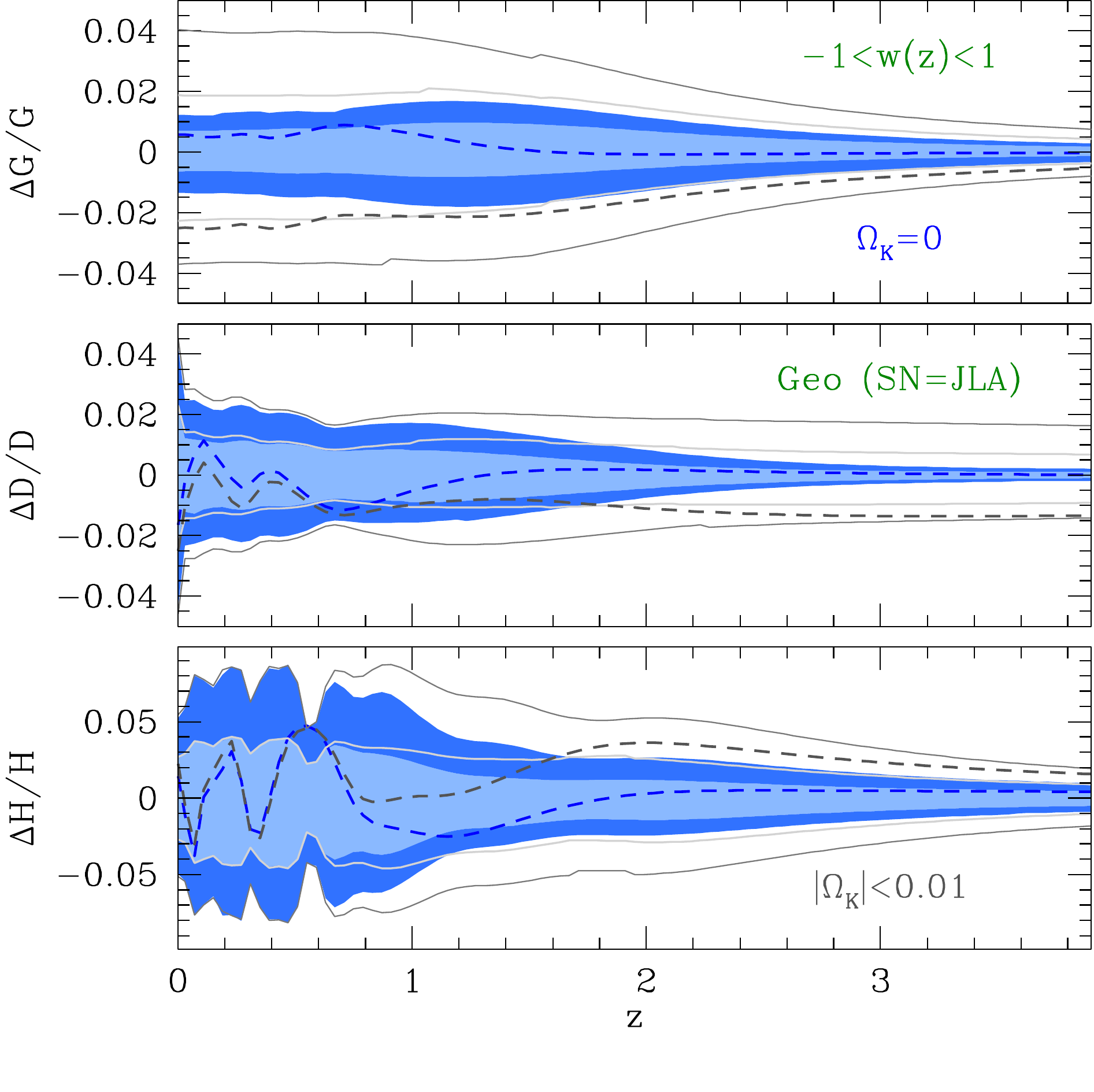}
\includegraphics[scale=0.44]{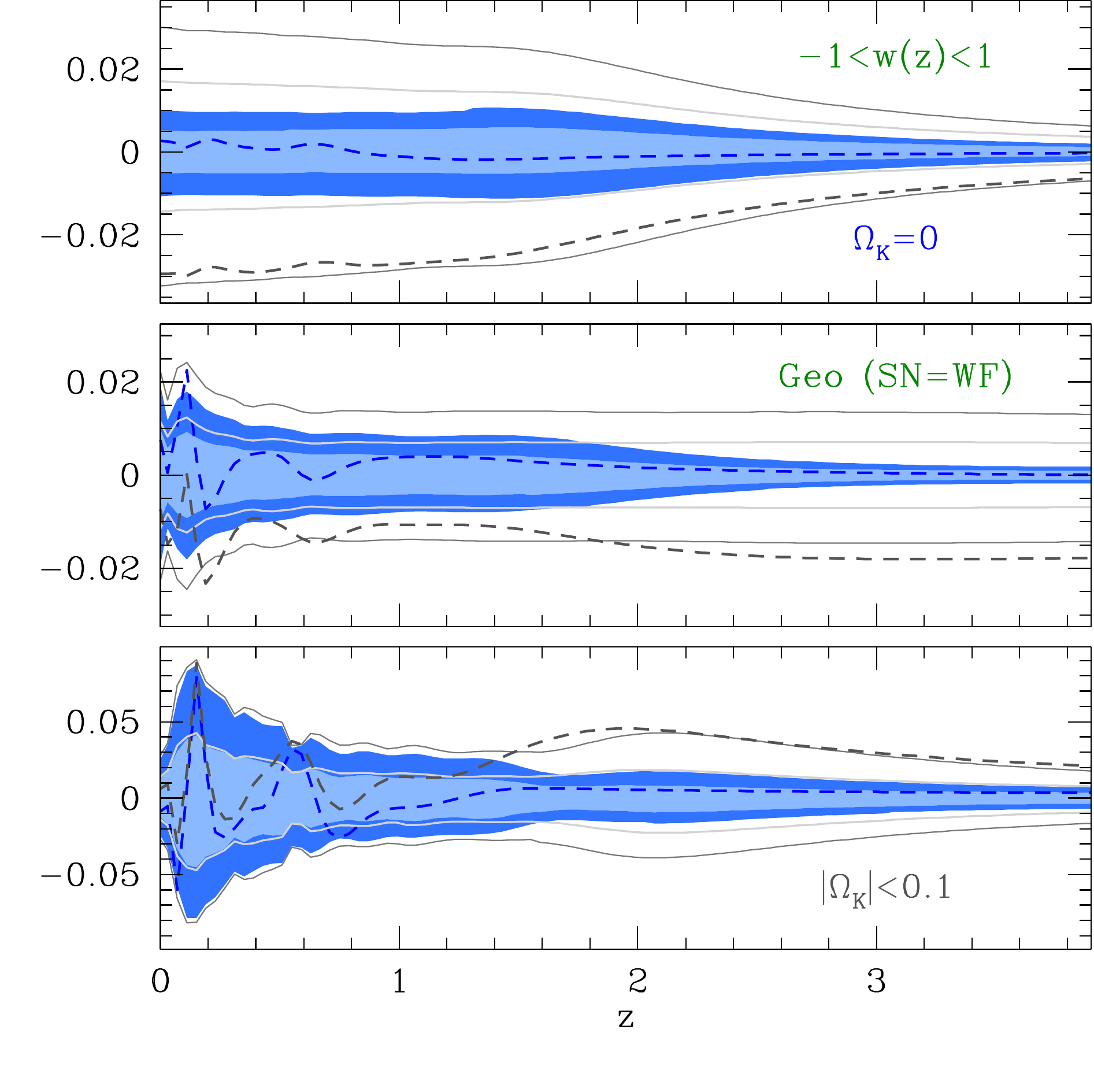}
\caption{Similar to Figure~\ref{fig:LCDM1}, but in the context of quintessence. For future WFIRST data, we relax the spatial curvature prior to $ |\Omega_k| < 0.1$. Current observations constrain the Hubble expansion rate better than $10\%$ in flat models, but predictions above $z>1$ depend considerably on the allowed curvature. The more stringent $|\Omega_k| < 0.01$ prior is necessary to obtain percent level predictions with current data (see section~\ref{sec:growth_probes} for further discussion on this issue). WFIRST, on the other hand, tightens the distance posterior to a few percent even when marginalizing over arbitrary curvature. Finally,  the growth function in quintessence never exceeds $\Lambda$CDM predictions by more than $2-3\%$, and this allows both models to be simultaneously falsified \cite{Mortonson:2008qy}.}
\label{fig:quintessence1} 
\end{figure*}

To ensure that the dark energy equation of state respects the prior $w_\text{min} < w(z) < w_\text{max}$, we follow the procedure described in Appendix A of Ref. \cite{Mortonson:2008qy}. We start from the projection of a generic $w(z)$ on the PC basis
\begin{align}
\alpha_i = \frac{1}{N_z} \sum_{j=1}^{N_z} \, [w(z_j) - w_\text{fiducial}]e_i(z_j).
\end{align}
Now, the maximum/minimum $\alpha_i$ values are achieved whenever $w(z_j) = w_\text{max}/w_\text{min}$ and $e_i(z_j)$ is positive, and $w(z_j) = w_\text{min}/w_\text{max}$ and $e_i(z_j)$ is negative. Therefore, we require that
$\alpha^{(-)} < \alpha < \alpha^{(+)}$, with
\begin{multline} \label{eqn:prior1}
\alpha^{(\pm)}_i \equiv \frac{1}{N_z} \sum_{j=1}^{N_z} \big[(w_\text{min} + w_\text{max} - 2w_\text{fiducial})e_i(z_j) \\\pm (w_\text{max}-w_\text{min})|e_i(z_j) | \big].
\end{multline}
Last, we further impose the prior on the sum
\begin{multline}
\sum_{i=1}^{N_z} \, [w(z_j) - w_\text{fiducial}]^2 <  \\ \sum_{i=1}^{N_z} \text{max}[(w_\text{max}-w_\text{fiducial})^2, (w_\text{min}-w_\text{fiducial})^2],
\end{multline}
which implies
\begin{align} \label{eqn:prior2}
\sum_{i=1}^{N_{\text{PC}}}  \alpha_i^2 < \text{max}[(w_\text{max}-w_\text{fiducial})^2, (w_\text{min}-w_\text{fiducial})^2]. 
\end{align}
These are conservative priors because not all equations of state that respect the inequalities in Eqs.~\ref{eqn:prior1}, and~\ref{eqn:prior2} are limited to the range $w_\text{min} < w(z) < w_\text{max}$, but the converse is true, i.e., the priors keep all the models we want and eliminate many models we need to exclude. 

\begin{figure*}[t]
\includegraphics[scale=0.44]{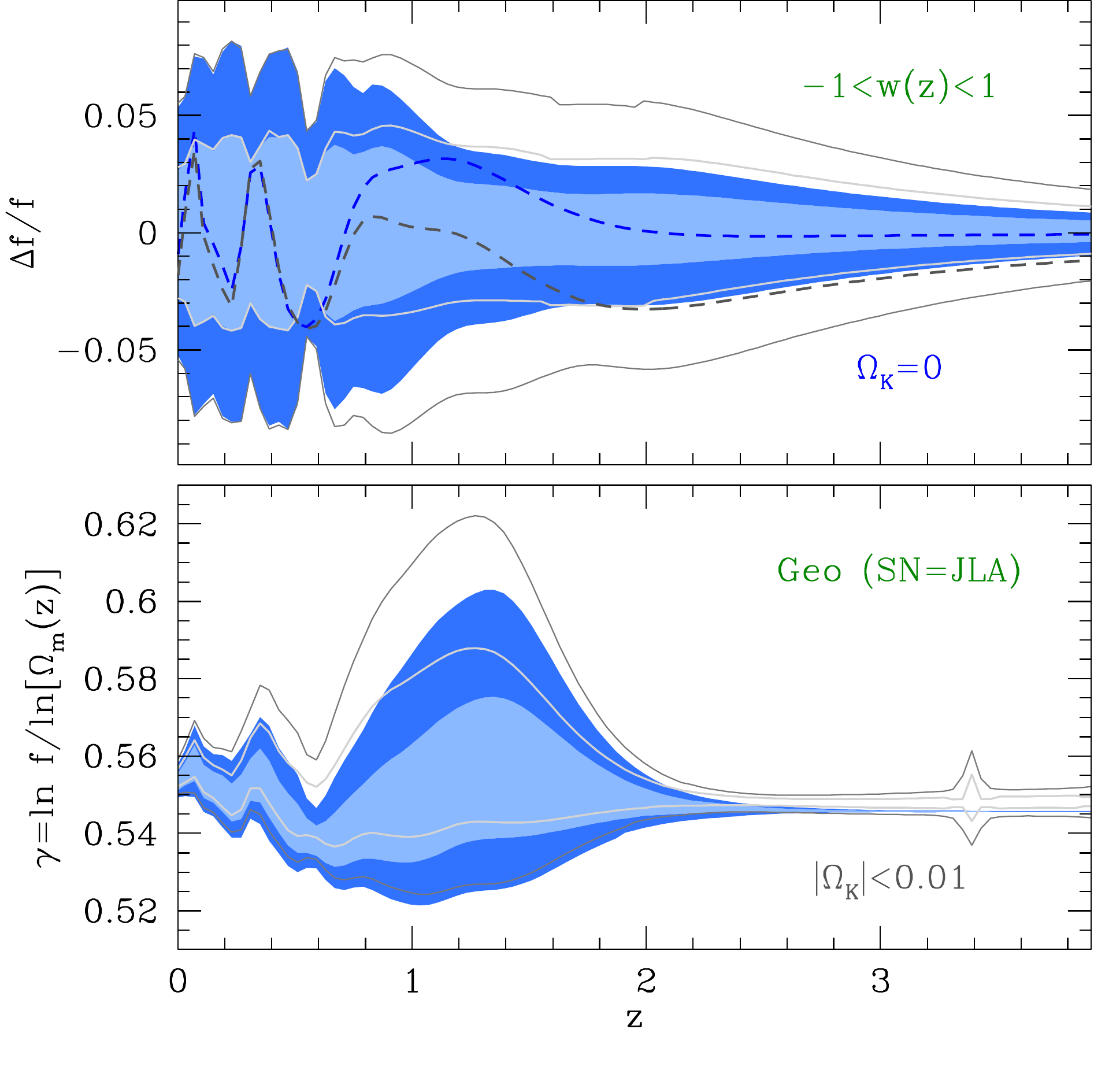}
\includegraphics[scale=0.44]{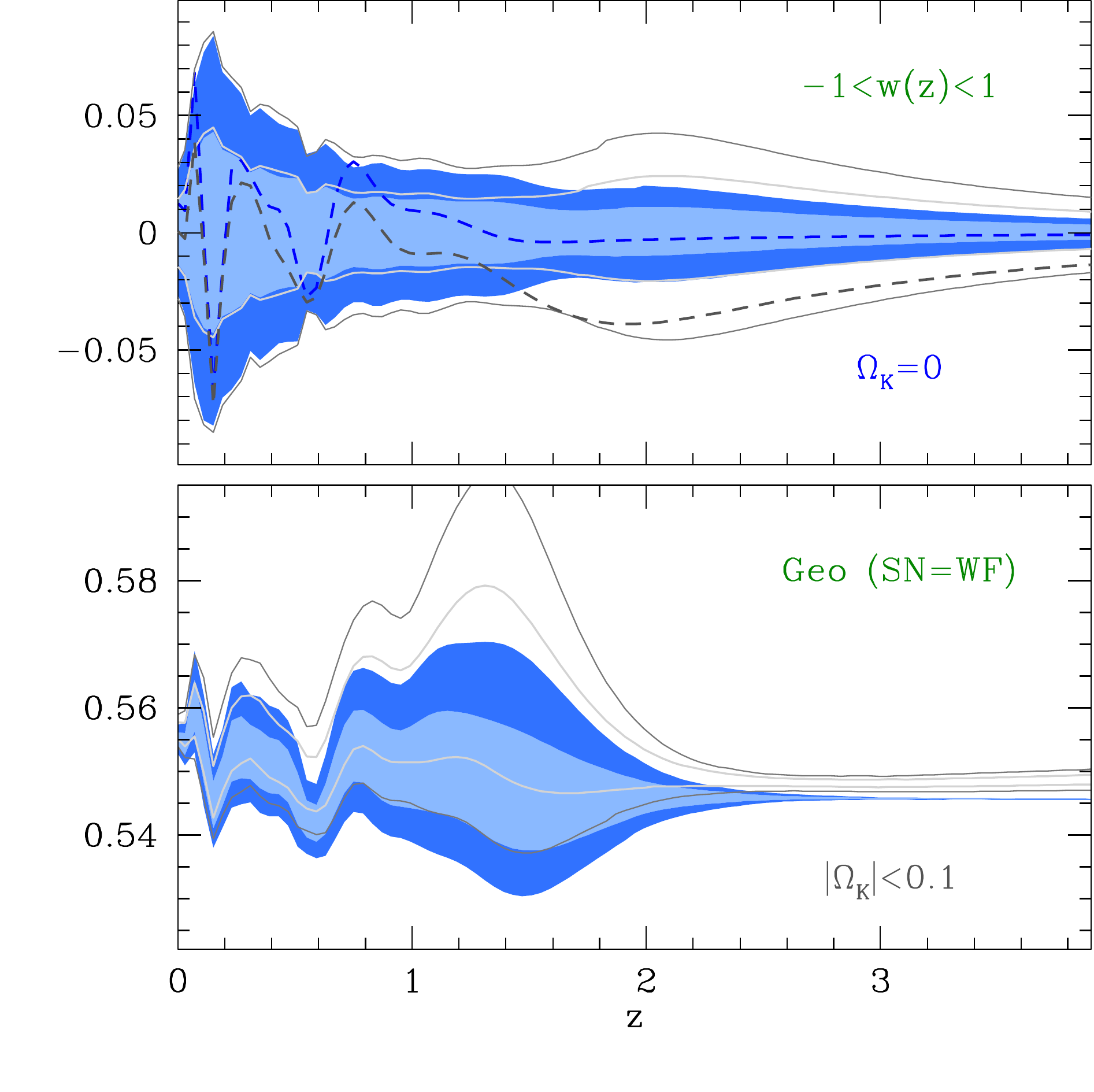}
\caption{Similar to Figure~\ref{fig:LCDM2}, but in the context of quintessence. For future WFIRST data, we relax the spatial curvature prior to $ |\Omega_k| < 0.1$. In all models, dark energy is modeled with 20 PCs at $z<z_\text{max}=3$ and with $w(z>z_\text{max})\equiv w_\infty = -1$.}
\label{fig:quintessence2} 
\end{figure*}


\section{Falsifying Smooth Dark Energy}
\label{sec:main_analysis}

In this section, we investigate the falsifiability of smooth dark energy scenarios. The observables used here only constrain the background expansion of the universe (Table~\ref{table:datasets_geo} describes them in further detail). The data include local $H_0$ measurement, baryon acoustic oscillations, comoving distance to the surface of the last scattering, and Type IA supernovae. They are collectively described as the {\it Geo} (SN=X) datasets, with $X = \text{JLA}$ or $X = \text{WF}$ (WFIRST) representing the adopted Type IA supernovae data. In some aspects, this section provides a partial update to the analyses presented on Refs.~\cite{Mortonson:2008qy,2010PhRvD..81f3007M}. The WFIRST supernovae data was simulated using state-of-the-art numerical tools, and the final likelihood takes into account a variety of systematic effects described in detail in Ref.~\cite{2017arXiv170201747H}.

Smooth dark energy models modify the amplitude of linear perturbations only through changes in the background evolution. Because the data contained in the {\it Geo} group are sensitive uniquely to the background expansion, the posteriors for the linear growth function are predictions that can be falsified with surveys that measure the amplitude of fluctuations. In fact, the Dark Energy Survey (DES) collaboration already released its Year One data that can potentially falsify the predictions presented in this section. The WFIRST satellite, on the other hand, will release its supernovae results only by the end of the next decade and, at that time, it will also provide state-of-the-art weak lensing measurements that can be used to check the consistency between growth and geometry in smooth dark energy scenarios.
 
We examine chains that assume flatness, $\Omega_K=0$, and others that allow spatial curvature to be a free parameter within some pre-specified width. The prior of $-0.01<\Omega_K< 0.01$ was adopted in chains with current type IA supernovae data, so predictions can be at the few percent level. This prior is, indeed, informative except for $\Lambda$CDM models. Percent level upper limits on $|\Omega_K|$ are at the order of what can be achieved by current data when growth information is included. For chains with simulated WFIRST Type IA supernovae, the prior width is relaxed to $-0.1<\Omega_K< 0.1$, which is not informative given the few percent level constraints we obtain in this case.


\subsection{$\Lambda$CDM}
\label{sec:FLCDM}

In flat $\Lambda$CDM, the geometric data set with current JLA supernovae can constrain the comoving luminosity distance at the sub-percent level at all redshifts, despite the fact that $H_0$ is measured at the $2.4\%$ level. The Hubble expansion rate shows a striking tightness around $z \approx 0.9$, which was previously noted in Ref.~\cite{Mortonson:2008qy}. With the CMB's outstanding precision in measuring the distance to the surface of last scattering, changes in the comoving distance at high redshift must be compensated by an opposite variation at low redshift where $ D(z) \sim z/H_0$. Changes in $H_0$ at the few percent level are, therefore, not compatible with the CMB and high redshift Type IA supernovae, even though they are allowed by local measurements. More precise local $H_0$ measurements would reduce uncertainties in predicting the comoving distance at high redshift in $\Lambda$CDM scenarios, which in turn would increase the ability of future Type IA supernovae missions to falsify the standard model with no spatial curvature. A related issue is a well-known tension between local $H_0$ measurements and the Hubble constant inferred from the CMB acoustic peaks. One could ask if this discrepancy is artificially tightening the constraints in flat $\Lambda$CDM. While it does impact the mean spatial curvature posteriors towards positive values, which also shifts the mean $H(z)$ predictions as seen in Figure~\ref{fig:LCDM1}, it does not affect the width of the posteriors significantly as we demonstrate in Figure~\ref{fig:test_H0}.  

Allowing curvature to be a free parameter degrades the $95\%$ contours on the growth function and the comoving angular diameter distance by a factor of $\approx 2$ (see Figure~\ref{fig:LCDM1}). Moreover, the posterior of the Hubble parameter becomes monotonic with redshift, i.e., there is no more degeneracy between changes in the $H_0$ at low redshift and variations in the comoving distance at high redshift. In the curved scenario, the posterior means of all the three functions shown in Figure~\ref{fig:LCDM1} shift at the two sigma level in comparison to the flat $\Lambda$CDM case. These variations are mainly induced by the discrepancy between the local $H_0$ measurements and the angular position of the CMB peaks. Indeed, a positive spatial curvature is correlated with higher $H_0$ prediction at a fixed angular position of the CMB peaks. 

Going from the current JLA data to the simulated WFIRST Type IA supernovae tightens posteriors by $40\%$ to $50\%$ in both flat and curved $\Lambda$CDM scenarios. The MCMC runs with simulated WFIRST supernovae are somewhat pessimistic because they assume that errors in $H_0$ will still be at the $2.4\%$ level by the end of the next decade while Gaia and JWST could potentially bring the errors down to a sub-percent level \cite{Riess:2016jrr}. Another possibility for measuring $H_0$ with better precision may come from strong-lensed type IA supernovae. However, refinements in the $H_0$ precision could also exacerbate the current tension between local measurements and the CMB. Even with a measurement of the Hubble constant at $2.4\%$, the middle panel in Figure~\ref{fig:test_H0} shows that disagreement between CMB and local $H_0$ measurements creates a systematic shift in the posterior mean at the $2$ sigma level.

\begin{figure}[t]
\vspace{1ex}
\includegraphics[scale=0.44]{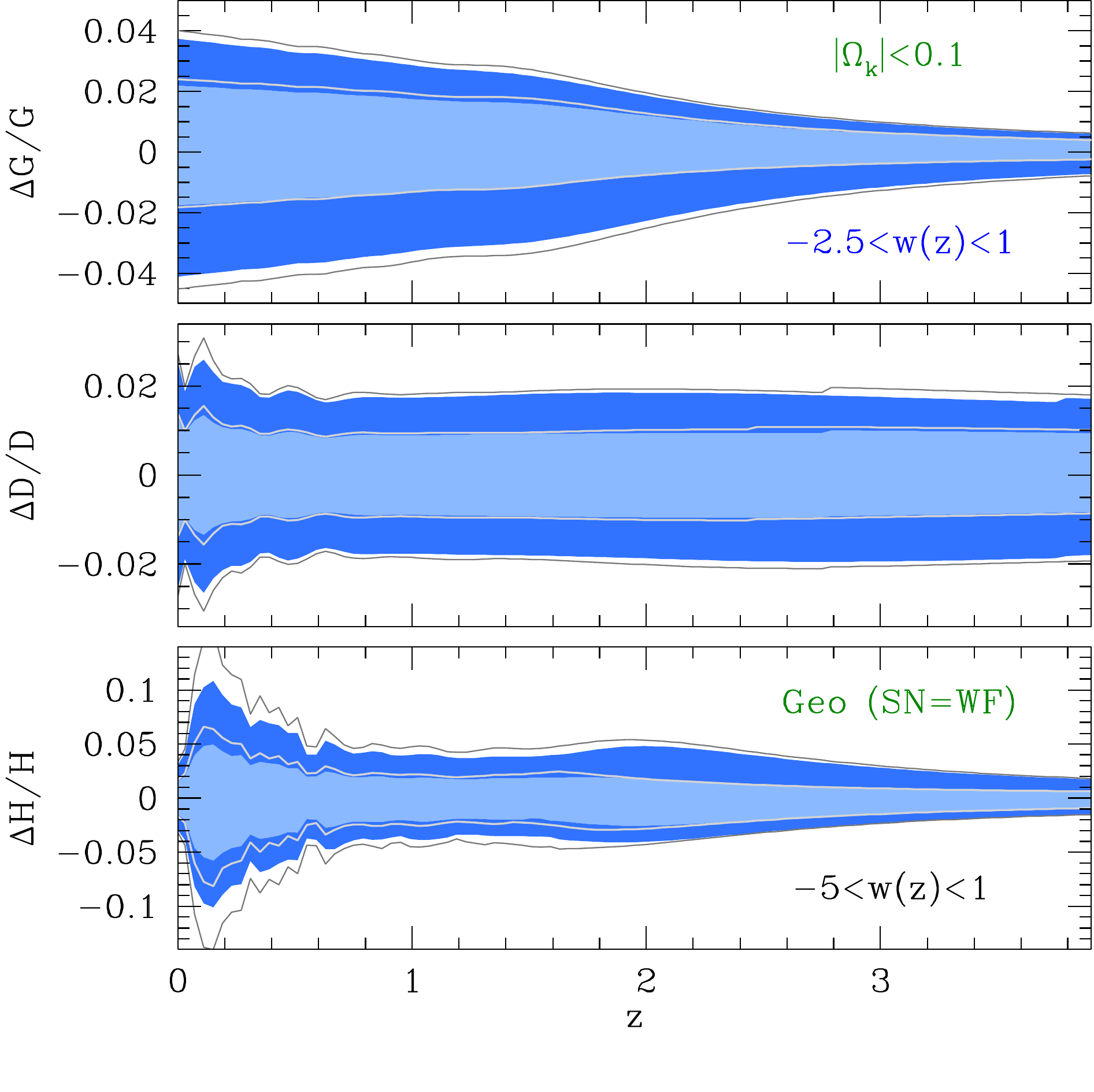}
\caption{This figure shows how the lower bound of the dark energy equation of state affects the growth function, the comoving luminosity distance, and Hubble expansion rate posteriors within the smooth dark energy scenario. Solid black lines display the $68\%$ and $95\%$ confidence levels with the prior $-5 < w(z) < 1$, while the blue shaded regions correspond to posteriors for the chain with $-2.5 < w(z) < 1$. In both cases, curvature is set to be a free parameter within the bounds $|\Omega_K| < 0.1$. Except at the lowest bins, differences in the distance and growth posteriors are at the percent level. The discrepancy in the Hubble in the Hubble expansion rate is significantly larger than in the distances, indicating that amplitude shifts of highly oscillatory modes are responsible for the change in the Hubble posterior.}
\label{fig:PCA_WFIRST_WZ_RANGE}
\end{figure}

Constraints on the growth function are approximately at the $0.5\%$ level with current data in flat $\Lambda$CDM. Marginalization over spatial curvature increases the error on the growth function by a factor of $2$, and it also shifts down the growth posterior mean by almost a percent at redshift $z=0$. The growth rate shows a similar behavior (see Figure~\ref{fig:LCDM2}). Future updates on the analysis presented in Ref. \cite{2011PhRvD..83b3015M}, which translates growth predictions to counts of massive clusters, may prove worthwhile to be pursued given that the current $H_0$ discrepancy pushes the growth function in curved models to values below the flat case. Finally, Figure~\ref{fig:LCDM2} shows that $\Lambda$CDM has tight predictions on the growth index, which offer an alternative test that can be used to falsify the standard model.


\subsection{Quintessence}
\label{sec:FQUIT}

Quintessence scenarios offer a wider range of predictions that could still be compatible with data even in the case in which the $\Lambda$CDM scenario is falsified by future surveys. From a physical standpoint, scalar fields with dynamics dictated by Lagrangians of the form $\mathcal{L} = X - V$, where X and $V$ are the field's kinetic energy and potential energy respectively, could drive the accelerated expansion with an equation of state that remains above the phantom barrier of $w = -1$. Quintessence is, therefore, one of the simplest $\Lambda$CDM generalizations. In this section, we model quintessence scenarios with the complete set of principal components. To impose the canonical scalar field boundaries $-1 < w(z) < 1$, we adopt the set of conservative priors on the PC amplitudes that are shown in Eqs.~\ref{eqn:prior1} and~\ref{eqn:prior2}, with $w_\text{min} = -1$ and $w_\text{max} = 1$.

Quintessence predictions for the growth function are $4$ times broader relative to the ones assuming $\Lambda$CDM. Also, there is a $2$\% shift downwards in the growth's posterior mean, present in runs with either current JLA supernovae or future WFIRST simulated data. This ensures that the growth function in quintessence never exceeds the mean $\Lambda$CDM expectation by more than approximately  $2$\% percent \cite{Mortonson:2008qy}. Consequently, the growth posteriors showed in Figure~\ref{fig:quintessence1} provide an exciting possibility of falsifying quintessence and $\Lambda$CDM simultaneously, especially if gravity does not follow the predictions from General Relativity in scales well-modeled by linear perturbation theory. Indeed, modifications of gravity often introduce new degrees of freedom, and they generically enhance the amplitude of linear perturbations well above $\Lambda$CDM predictions. These type of theories generally have screening mechanisms to ensure compatibility between alternative models of gravity and stringent observations in our Solar System, and they show deviations from General Relativity at scales well-modeled by linear perturbation theory. 

\begin{figure*}[t]
\includegraphics[scale=0.44]{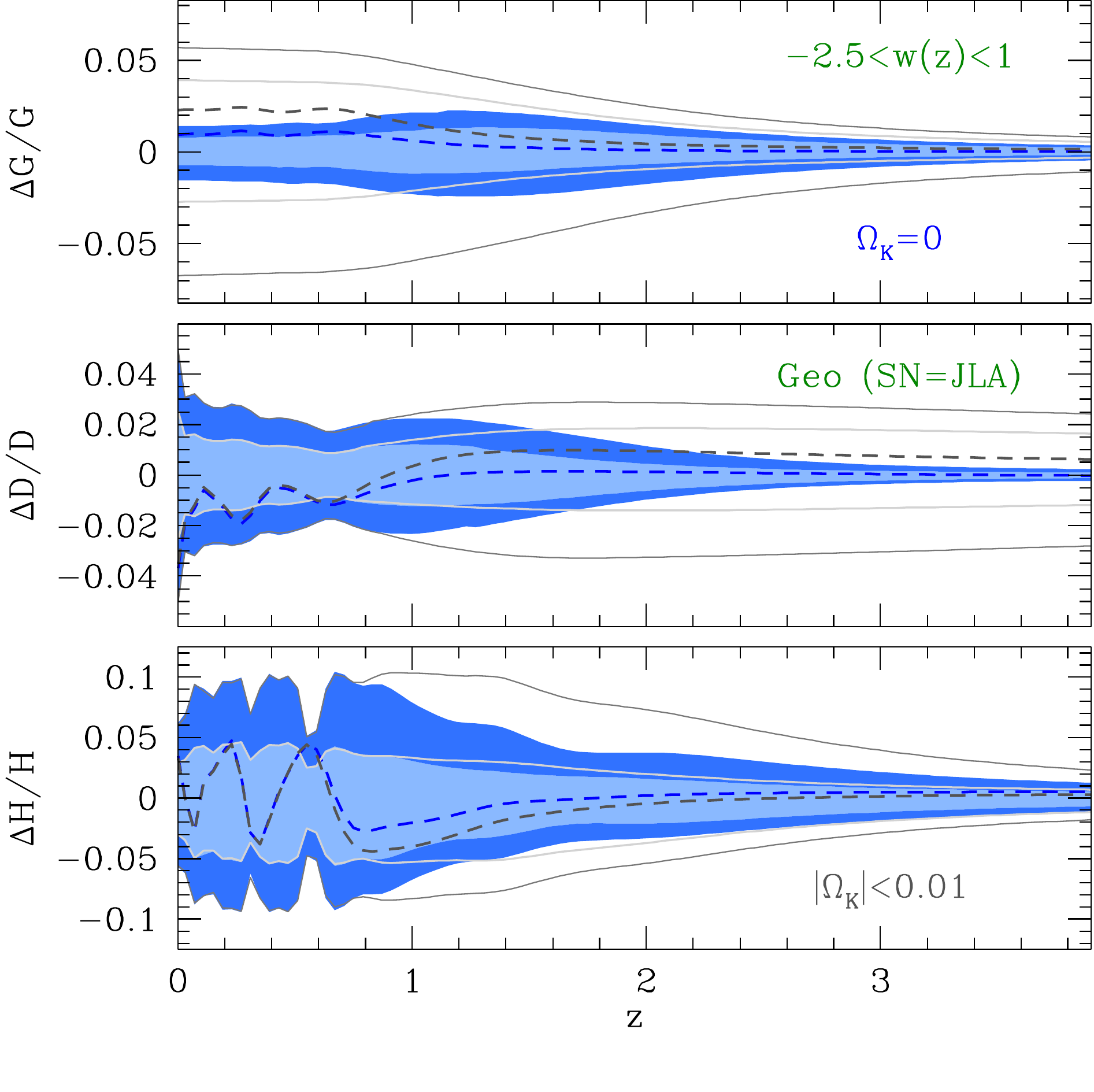}
\includegraphics[scale=0.44]{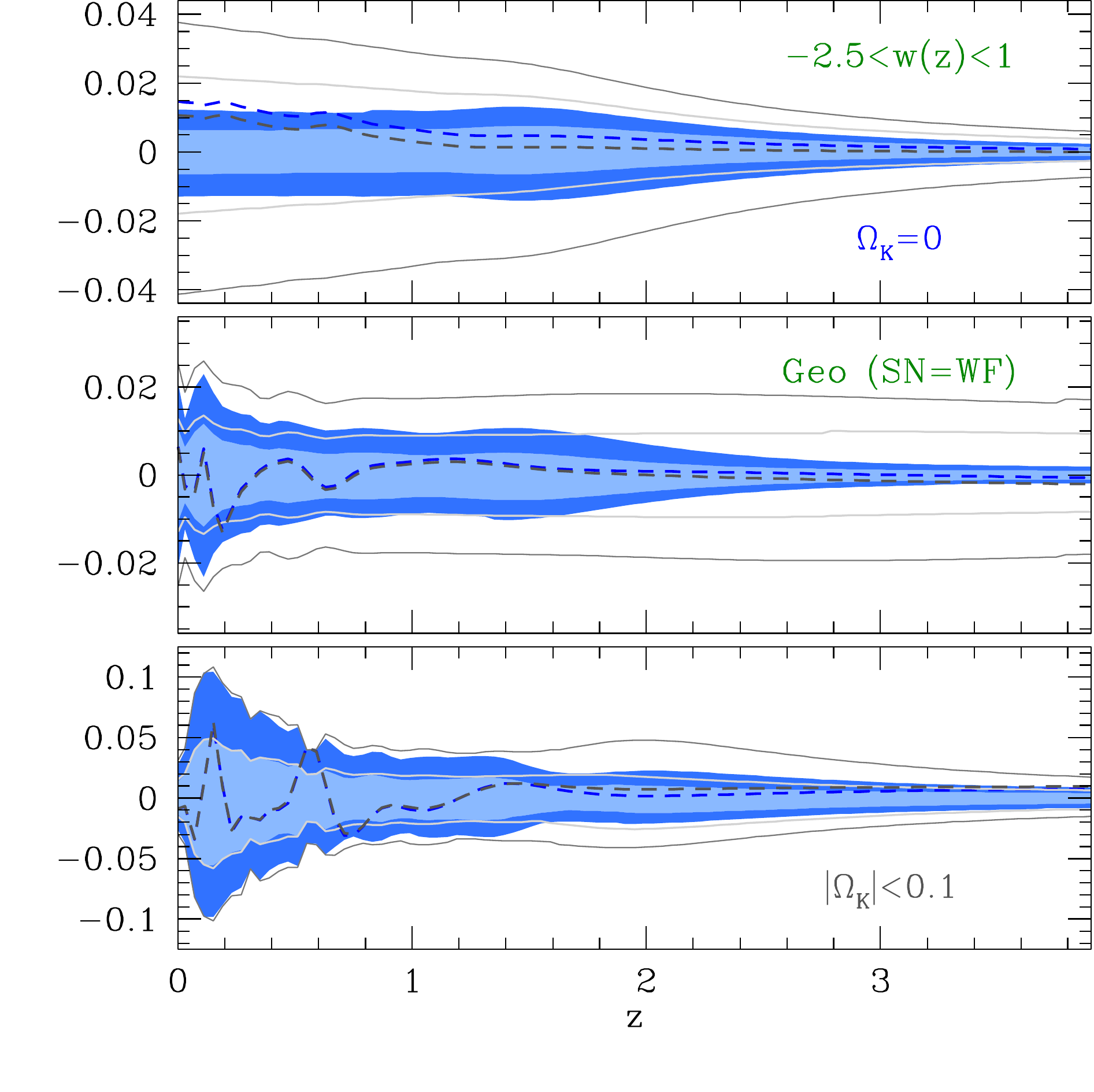}
\caption{Similar to Figure~\ref{fig:quintessence1}, but in the context of smooth dark energy scenarios, where the dark energy equation of state is restricted to the range $-2.5 < w(z) < 1$. The extra freedom provided by the phantom crossing in curved models widens the growth function posterior at redshift zero by approximately $40\%$ with current JLA supernovae and $25\%$ with WFIRST simulated data, in comparison to quintessence models (see Figure~\ref{fig:quintessence1}). General smooth dark energy scenarios also predict values for the growth function that are substantially higher ($5\%-7\%$) than the $\Lambda$CDM predictions. Future WFIRST should be able to constrain the comoving distance at the $2\%$ level and the growth function at the $4\%$ level even when marginalizing over arbitrary curvature.}
\label{fig:smooth} 
\end{figure*}

The comoving luminosity distance posteriors in flat quintessence models are about twice as wide as in $\Lambda$CDM. At the same time, the Hubble function is a factor five as broad as the standard model at the redshift range $0 < z < 1$. The broadening of the Hubble function is mainly due to highly oscillatory and not well constrained modes that are suppressed in the comoving luminosity distance. Indeed, the integration of the Hubble function smooths oscillatory behavior. In flat quintessence, chains with WFIRST type IA supernovae show a  $30\%$ improvement in precision in comparison to that obtained with current data. On the other hand, WFIRST constraints show order unity improvements when spatial curvature is a free parameter. Indeed, present data is not powerful enough to provide percent-level predictions when marginalized over the more extensive range $|\Omega_K| < 0.1$. WFIRST, on the other hand, will be able to constrain the comoving luminosity distance at the $3\%$ level, even when marginalized over arbitrary values of $\Omega_K$. Finally, both flat and curved quintessence scenarios show an order of magnitude broadening in the growth index posterior in comparison to $\Lambda$CDM. The growth rate posterior is also wider, by a factor of $2$ approximately, in contrast to $\Lambda$CDM (see Figure~\ref{fig:quintessence2}). 


\subsection{Smooth Dark Energy}
\label{sec:FSDE}

In this subsection, we have adopted the prior $-2.5 < w(z) < 1$ on the dark energy equation of state to reduce the computational requirements of the demanding MCMC likelihood analysis that we present here and in the subsequent sections. To quantify the loss of generality, Figure ~\ref{fig:PCA_WFIRST_WZ_RANGE} compares the result from runs with simulated WFIRST data and free curvature, where we assume either $-2.5 < w(z) < 1$ or $-5 < w(z) < 1$. The change in the $w(z)$ prior widens the growth function at redshift $z=0$ by no more than $15\%$. Nonetheless, an even ampler range in the equation of state together with the possibility that dark energy could have been relevant at earlier times may degrade growth predictions by a considerable amount. Also, the lack of constraining power in the spatial curvature stretches the posteriors in runs with current data by more than $15\%$. In any case, falsifying smooth dark energy models with $-2.5 < w(z) < 1$ and no significant amount of early dark energy would already be an enormous step towards motivating more exotic dark energy scenarios.   

In comparison to quintessence, Figure~\ref{fig:smooth} shows that crossing the phantom barrier widens the growth function posterior at $z=0$ by $40\%$ and the comoving luminosity distance posterior  by $30\%$ at $z>2$, in MCMC runs where we use current data and $|\Omega_K|<0.01$ prior on the spatial curvature. With WFIRST simulated data and $|\Omega_k| < 0.1$, predictions for the growth function at redshift $z=0$ are at least $50\%$ larger, while for the comoving luminosity distance they are about $25\%$ broader, relative to the quintessence scenario. Both the growth function and the comoving luminosity distance posterior means are displaced by a few percent in comparison to quintessence predictions. Unlike in quintessence scenarios, the growth function in smooth dark energy models can exceed flat $\Lambda$CDM predictions by more than $2$\%. 

Figure~\ref{fig:curvature_geo_future} shows that the geometric data with simulated WFIRST supernovae will be able to probe the spatial curvature at the percent level. Indeed, WFIRST will be able to constrain the dark energy dynamics so tightly that there will not be enough freedom to compensate the shifts in the comoving distance induced by changes in curvature to maintain the distance to the last scattering fixed. With current data, however, constraints on $\Omega_K$ are considerably relaxed, which motivates the analysis we present in Section~\ref{sec:growth_probes}. There, we show how the combination $\sigma_8 \Omega_m^{1/2}$ can break the degeneracy between the dark energy equation of state and the spatial curvature. The inclusion of weak lensing, CMB lensing reconstruction, and redshift space distortion data makes order unity difference in the dark energy Figure of Merit after marginalization over spatial curvature.

\begin{figure}[t]
\includegraphics[scale=0.44]{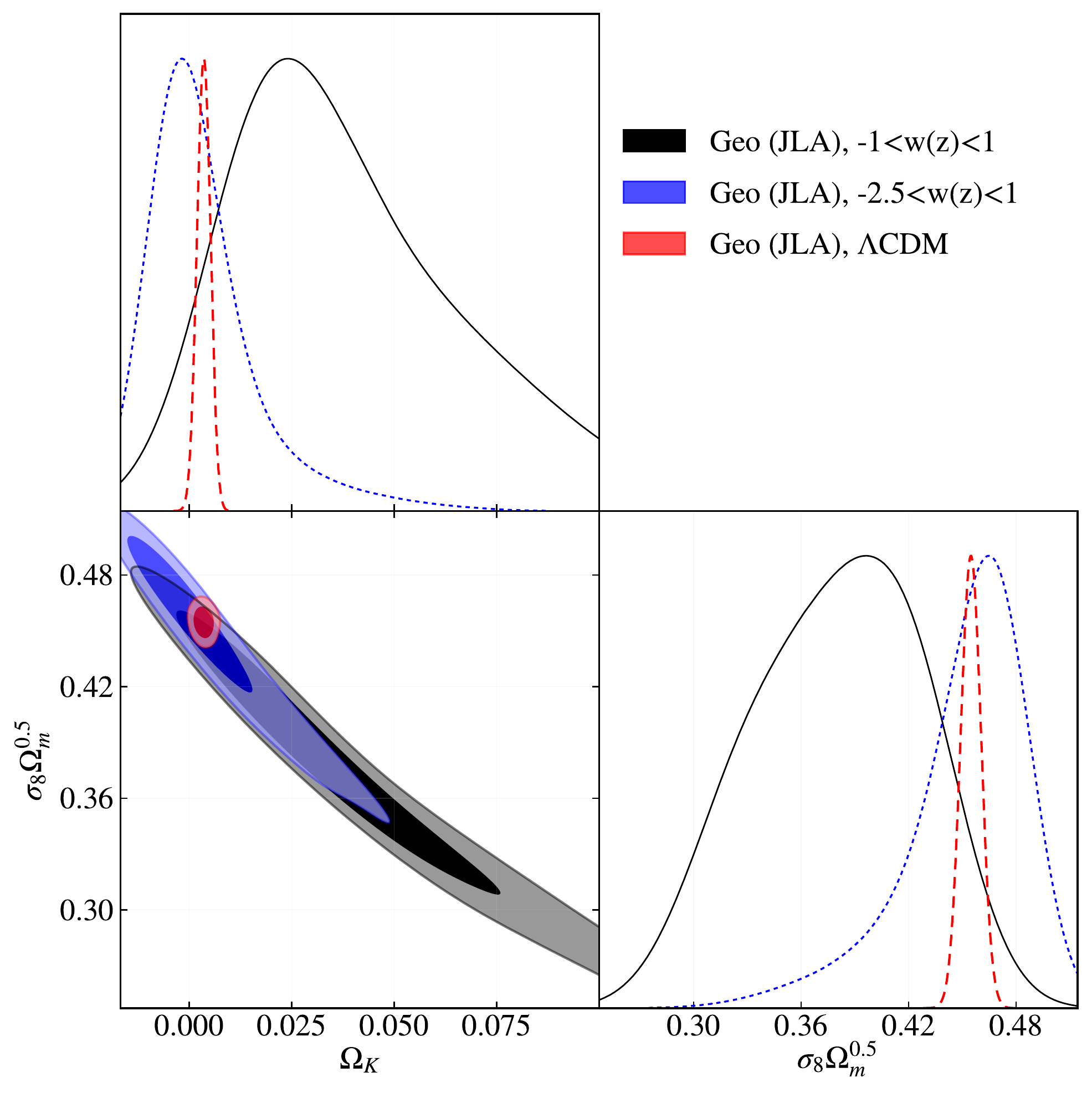}
\caption{Spatial curvature posterior and its strong correlation with the combination $\sigma_8 \Omega_m^{1/2}$. The inability of current geometric data to constrain spatial curvature at the percent level in both quintessence and general dark energy models provides an excellent opportunity for observables that directly measure the growth of structures to make a significant impact on the dark energy Figure of Merit. Finally, given that flat priors (that can also cross the phantom barrier) on the PC amplitudes translate into a preference for negative curvature, the reduction of the posterior probability for large and positive $\Omega_k$ in general dark energy models might be partially prior induced (see Figure 12 of~\cite{Mortonson:2008qy}).}
\label{fig:curvature_geo_current}
\end{figure}

\begin{figure}[t]
\includegraphics[scale=0.44]{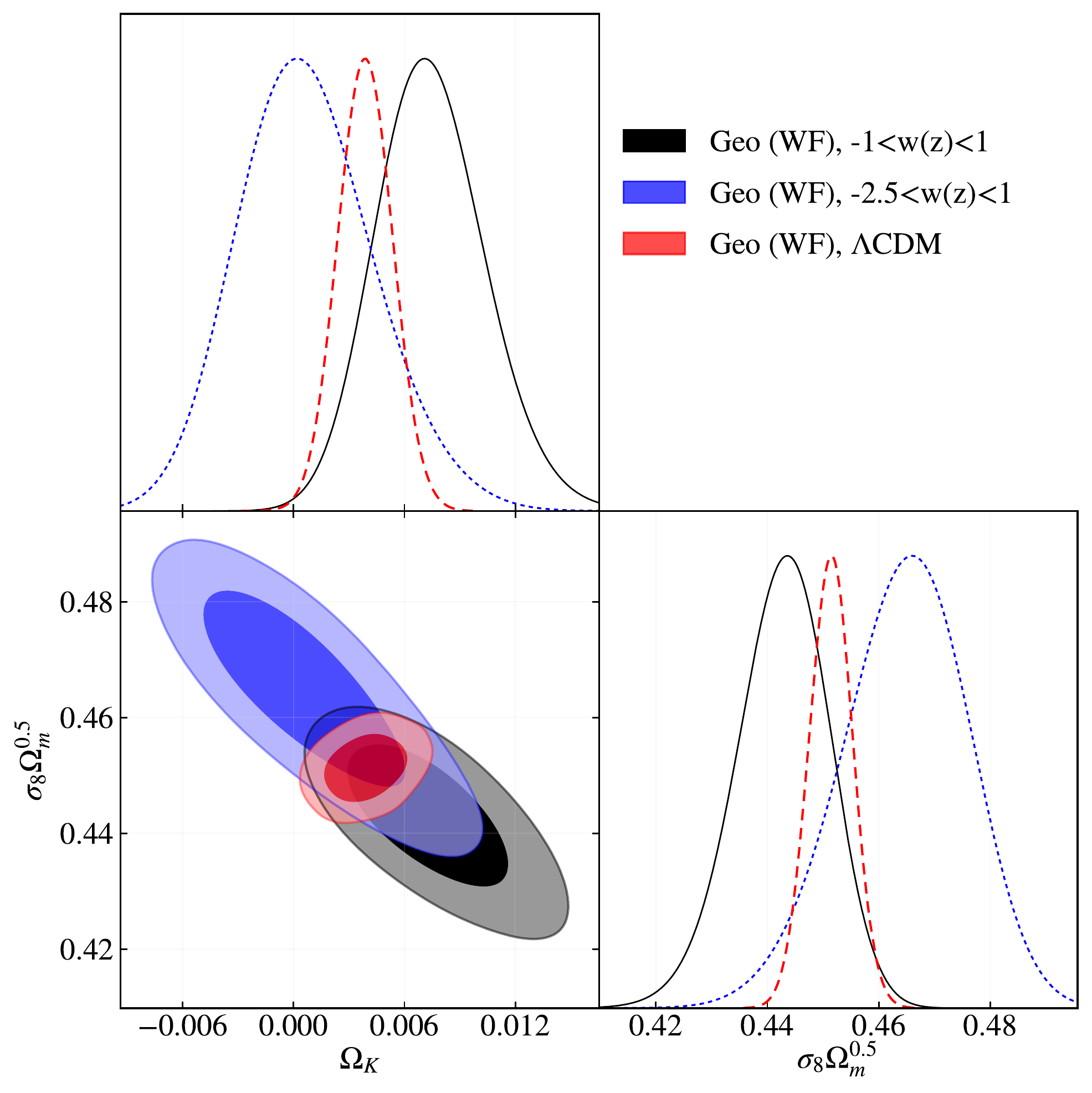}
\caption{Spatial curvature posterior and its correlation with the combination $\sigma_8 \Omega_m^{1/2}$. Further improvements in BAO measurements from the future DESI survey as well as advances in local $H_0$ measurements may bring $\Omega_K$ constraints to sub-percent level without growth information.}
\label{fig:curvature_geo_future}
\end{figure}

\begin{figure*}[t]
\includegraphics[scale=0.44]{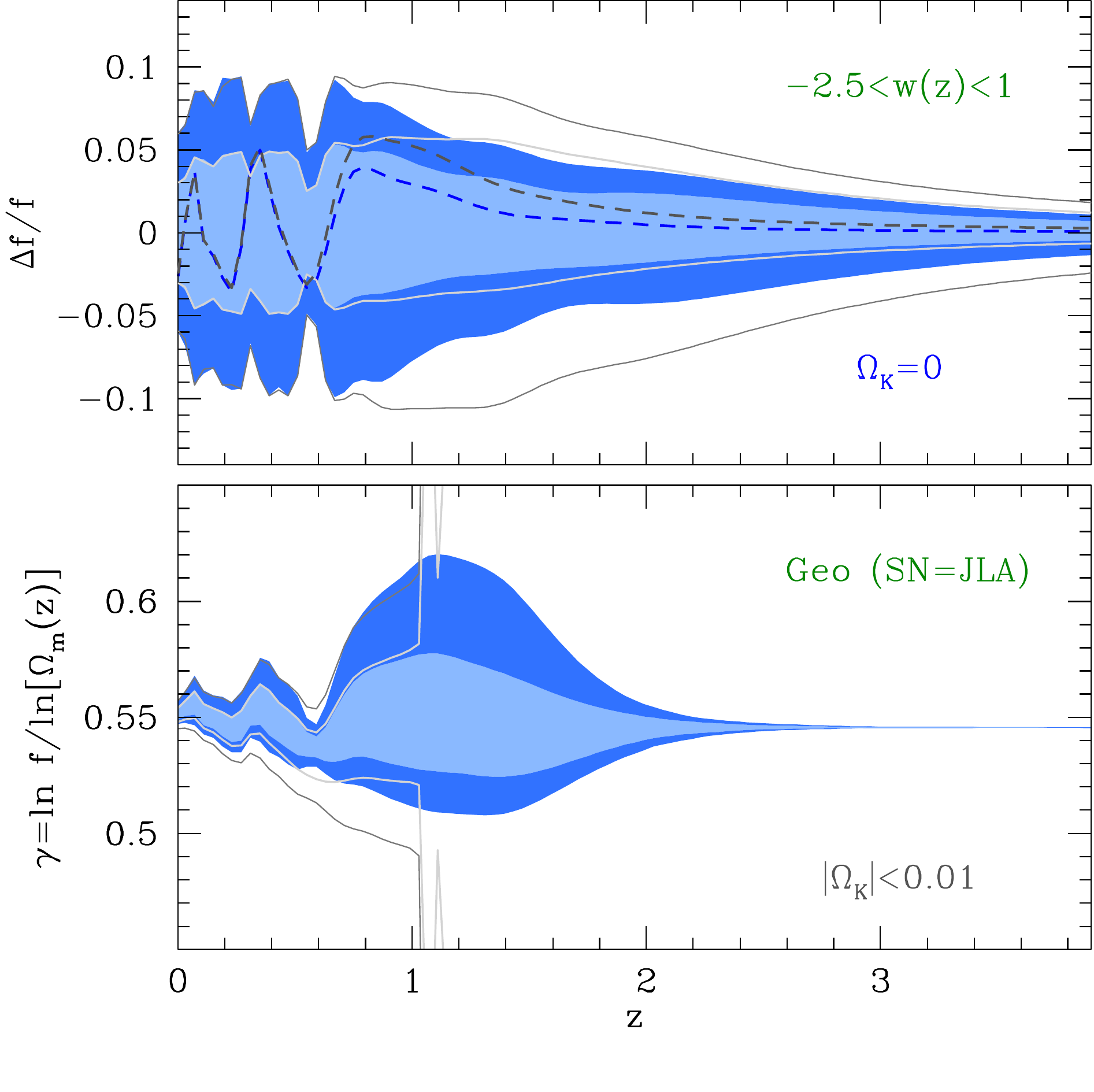}
\includegraphics[scale=0.44]{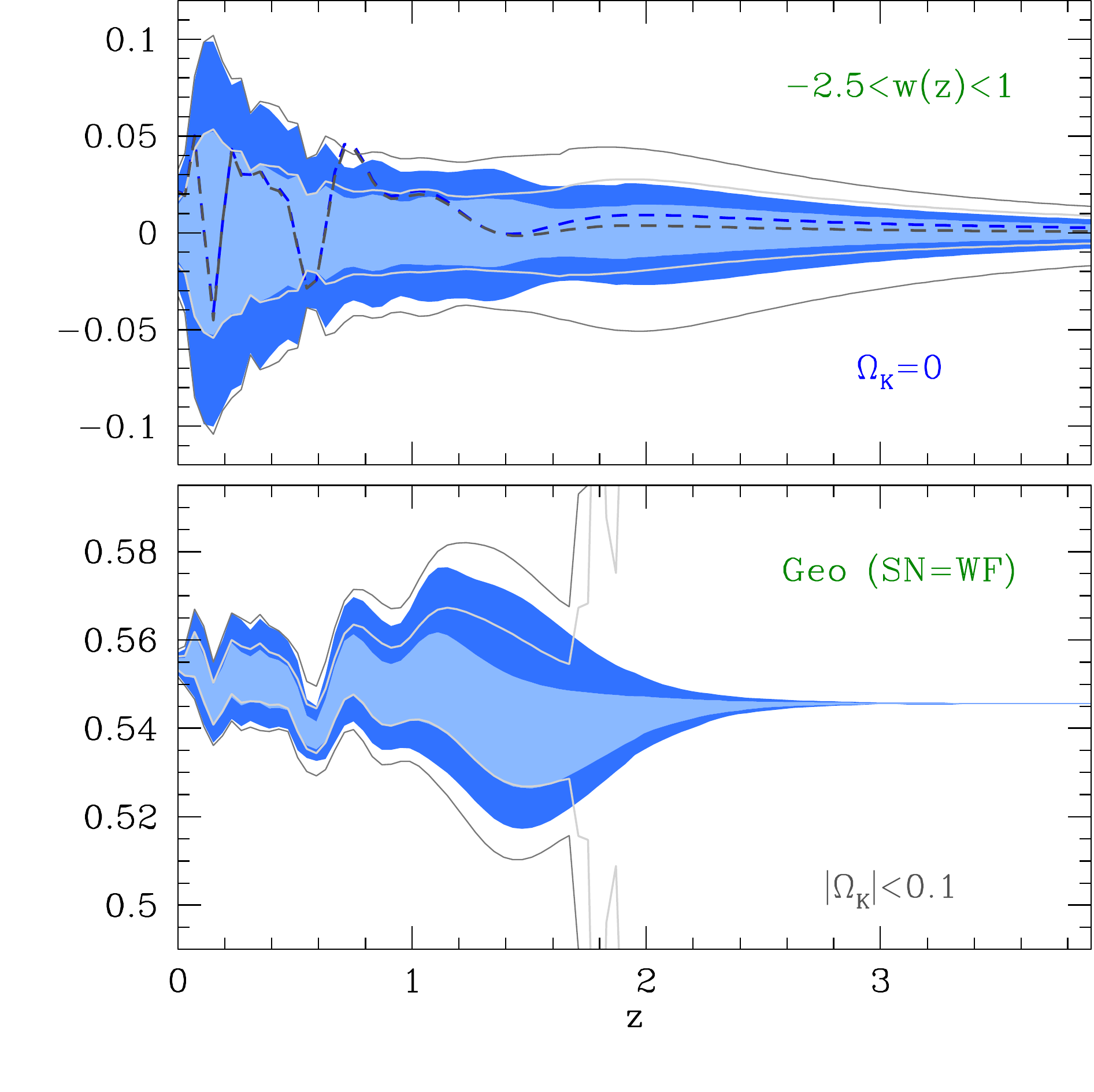}
\caption{Similar to Figure~\ref{fig:quintessence2}, but in the context of smooth dark energy models where the dark energy equation of state is restricted to the range $-2.5 < w(z) < 1$. The extra freedom provided by the phantom crossing in curved scenarios widens the growth rate by approximately $25\%$ above $z=1$. The posteriors for the growth index $\gamma(z) = \ln f(z)/\ln \Omega_m(z)$  become unstable immediately above redshift $z=1$. This problem in the growth index happens because $\Omega_m(z)$ can cross the boundary $\Omega_m(z)=1$ at high redshifts in curved scenarios. The same goes for $f(z)$, but the crossing $f(z)=1$ happens at slightly different redshifts \cite{Mortonson:2008qy}.  This unmatched crossing makes the posteriors to either change sign or to diverge, and hence the growth index loses its capability to falsify curved smooth dark energy scenarios.}
\label{fig:smooth2} 
\end{figure*}

Finally, Figure~\ref{fig:smooth2} shows that the posteriors for the growth index $\gamma(z) = \ln f(z)/\ln \Omega_m(z)$  become unstable immediately above redshift $z=1$. Even with the WFIRST simulated supernovae data, the growth index posteriors become ill-behaved above redshift $z \approx 1.5$. This problem in the growth index happens because $\Omega_m(z)$ can cross the boundary $\Omega_m(z)=1$ at high redshifts, in curved scenarios. The same goes for $f(z)$, but the crossing $f(z)=1$ happens at slightly different redshifts \cite{Mortonson:2008qy}. This unmatched crossing makes the posteriors either change signs or diverge, and hence, the growth index loses its capability to falsify curved smooth dark energy scenarios. The growth rate, on the other hand, is still well-behaved on the entire redshift range, and above $z=1$, its posteriors are about $25\%$ broader in smooth dark energy scenarios compared to quintessence models. This widening applies to chains with the current JLA data and with $|\Omega_K| < 0.01$ prior as well as to chains with the WFIRST simulated data and with  $|\Omega_K| < 0.1$. 

\section{Figure of Merit}
\label{sec:FOM}

In this section, we construct a model-independent definition for the Figure of Merit (FoM), making use of the principal components, following closely Ref. \cite{2010PhRvD..82f3004M}. We compute the FoM of the three cosmic acceleration scenarios we have studied so far: $\Lambda$CDM, quintessence, and general smooth dark energy scenarios where $-2.5<w(z)<1$. PCA-based FoM provides a complementary view to studies that assume particular functional forms for $w(z)$. These studies have the advantage of being more computationally efficient given the low number of parameters involved in parametrizations that are common in the literature. FoMs based on particular functional forms for $w(z)$ also have a straightforward  interpretation regarding signal-to-noise ratio. However, FoMs based on particular forms of $w(z)$ may underestimate or overestimate the constraining power of a given experiment, given that there are multiple compelling generalizations of $\Lambda$CDM, without a clear hierarchy between them in terms of theoretical plausibility\footnote{Indeed, non-parametric methods, such as PCAs are well suited to observables that cannot be robustly modeled from first-principle calculations. They have been used, for example, to describe inflation and the epoch of reionization~\cite{Dvorkin:2010dn,Heinrich:2016ojb}.}. They provide, therefore, an incomplete picture about the future capabilities of WFIRST in constraining dark energy models that predict more elaborate forms of $w(z)$. This incompleteness depends on how typical values for the amplitudes $\alpha_i$ compares with the $68\%$ and $95\%$ observational confidence levels on $\alpha_i$, when the fiducial model is projected on the PCA basis. This ratio is defined as the signal-to-noise ratio (see Equation 3 in Ref.~\cite{dePutter:2008bh}).

For example, Ref.~\cite{2010PhRvD..82f3004M} confirms that models in which the dark energy dynamics is dictated by a canonical field, $\phi$, that rolls on a potential of the form $V(\phi) = V_0 + m^2 \phi^2/2$ have small projected volumes in the subspace spanned by all except for the two most constraining principal components\footnote{The PCs of Ref. \cite{2010PhRvD..82f3004M} were constructed to mimic the discontinued SNAP experiment~\cite{Aldering:2004ak}.}. Indeed, Figure 5 of  Ref.~\cite{2010PhRvD..82f3004M} explicitly shows, for a particular choice of parameters, that only the first and second principal components have amplitudes that are comparable to their respective posterior uncertainties. While $V(\phi) = V_0 + m^2 \phi^2/2$ is a perfectly reasonable potential, there is not enough theoretical guidance from a more fundamental particle description of the dark energy component that prevent us from constructing more convoluted potentials that result in a $w(z)$ that needs to be described with more principal components. 

These nuances in interpreting the FoMs based on particular parametrizations, when there are multiple compelling dark energy models, can affect the design choices for future experiments in ways that could potentially reduce the possibility of discovering groundbreaking results. For example, the FoM based on $w(z) = \text{constant}$ models predict that the best supernovae strategies are the ones that focus their statistical power at the low redshift range $z < 1$. Similar conclusion can be derived from simple models where $w(z)$ is well described, in terms of signal-to-noise, by the first few PCA components of the {\texttt Imaging-Allz} strategy. However, observational strategies that focus on low-z supernova could lose the possibility of investigating models that predict $w(z)$ with large projected volume on the subspace spanned by higher principal components of the {\texttt Imaging-Allz} strategy (which would boost the signal-to-noise ratio of these components).

\begin{figure}[t]
\includegraphics[scale=0.44]{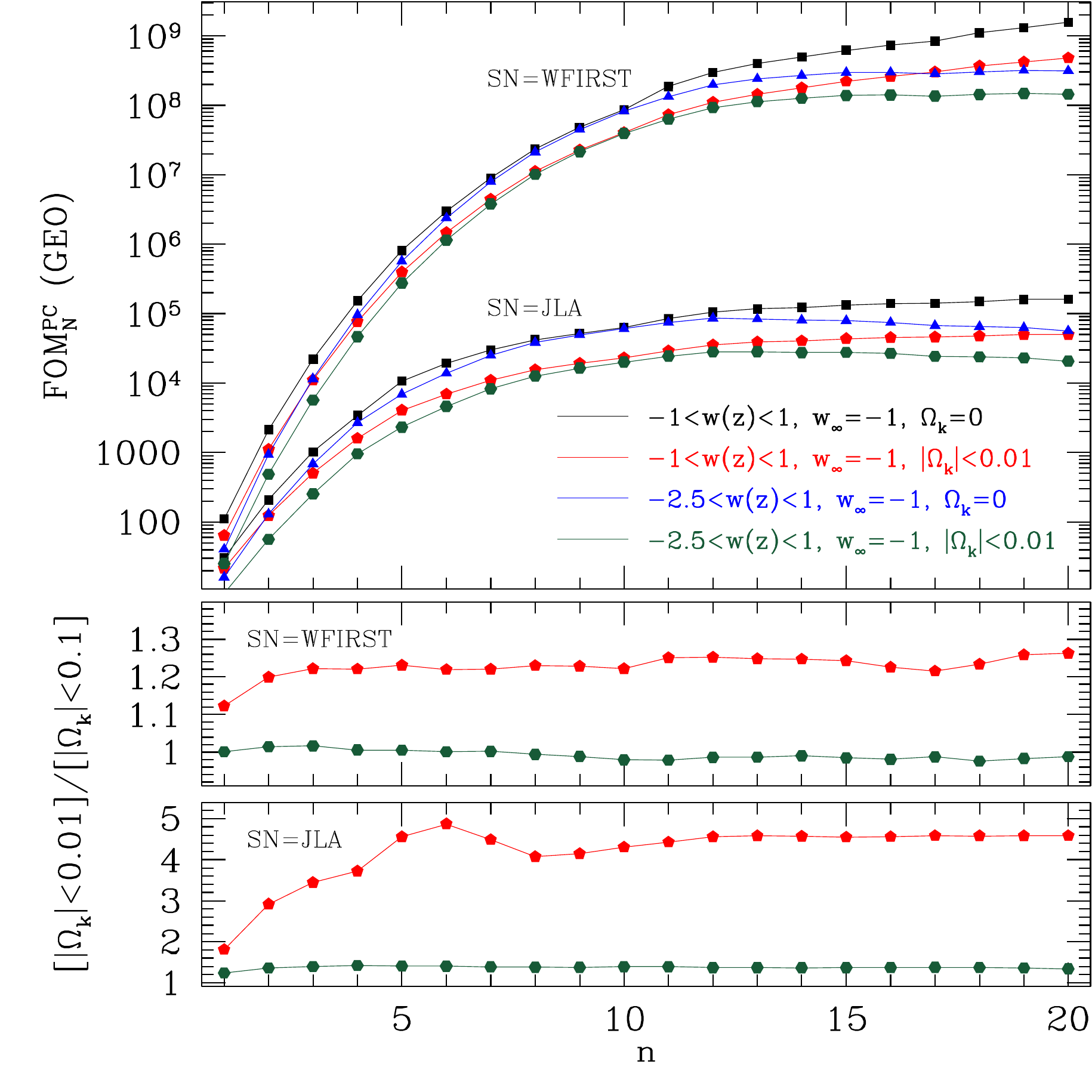}
\caption{Figure of merit of the PC amplitudes when the smooth dark energy scenario is probed with Planck geometric data, BAO, local $H_0$ and JLA/WFIRST Type IA supernovae measurements, as a function of the number of principal components. Current data can measure the first five principal components approximately. WFIRST will be able to measure these first five components significantly better and, in total, it will be able to constraint twice the number of modes. Finally, spatial curvature impacts the FoM derived from current data at order unity, while only at a few tens of percent with WFIRST simulated data.}
\label{fig:pca_fom_geo}
\end{figure}

Following Ref. \cite{2010PhRvD..82f3004M}, we define the Figure of Merit, given the covariance $\matr{C}_n$ between the principal components $e_i(z)$ with $i=1,...,n$, as
\begin{align} 
\text{FoM}_n^{\text{PC}} = \Bigg(\frac{\text{det} \,\, \matr{C}_n}{\text{det} \,\, \matr{C}_n^{\text{prior}}}\Bigg)^{-1/2}. 
\end{align}
Here, $\matr{C}_n^\text{prior}$ is the covariance of the prior, which we estimate based on MCMC chains that only takes into account the prior constraints on $w(z)$ in Eqs. \ref{eqn:prior1} and \ref{eqn:prior2}\footnote{This FoM definition depends on the prior volume, which might seem contrived, but by doing so we eliminate information gain that comes exclusively from the prior~\cite{2010PhRvD..82f3004M}. This can also be achieved by only considering the FoM ratio between two experiments~\cite{2009arXiv0901.0721A}.}. While not all parametrizations with $n$ parameters necessarily show the improvement given by $\text{FoM}_n$, this quantity represents an approximate upper limit of what is achievable with a given experiment. 
The quantity $\text{FoM}_n^\text{PC}$ is only an approximate upper limit because our PCs were developed with the Fisher matrix and not with the actual likelihood of the experiment. The interpretation of $\text{FoM}_n^\text{PC}$ with $n\ll 20$ as an upper limit for the background data with the current JLA data has an additional caveat, given that we use the same PC basis for both JLA and WFIRST experiments. However, the asymptotic value, $\text{FoM}_{n \approx 20}^\text{PC}$, can robustly be interpreted as an upper limit to both JLA and WFIRST because these $20$ PCs span a complete basis to both experiments. 

\begin{figure}[t]
\includegraphics[scale=0.44]{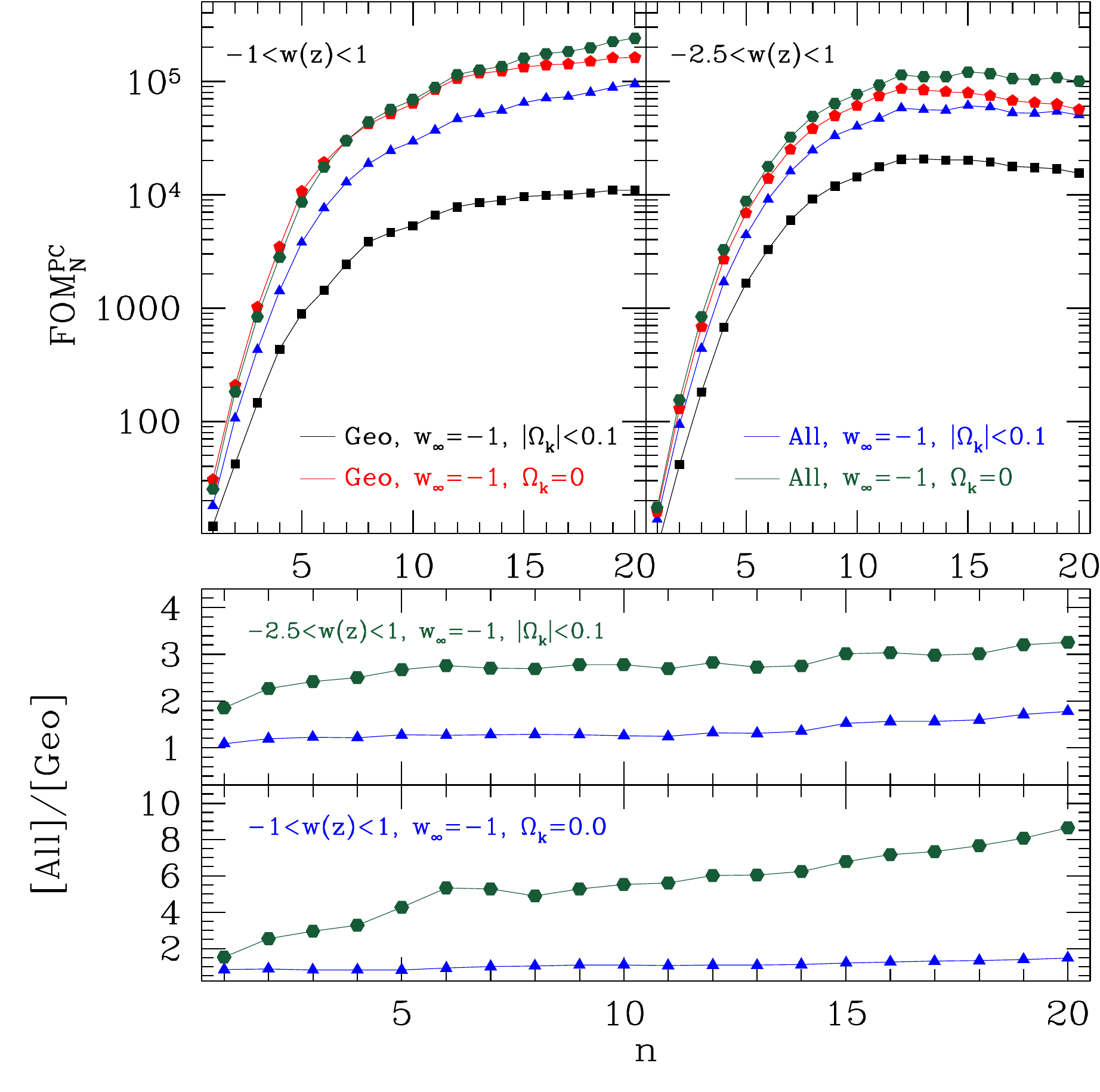}
\caption{The Figure of Merit of the PC amplitudes when quintessence (top left panel) and smooth dark energy (top right panel) are probed with either the {\it Geometric} or the {\it All} datasets. For quintessence, direct measurements of the growth of structure make an order of magnitude change in the FoM values, as it decreases the posterior probabilities of large and positive $\Omega_K$. For general smooth dark energy, the difference in the FoM is not as dramatic when growth information is combined, but given that flat priors to the PC amplitudes translate into a preference for negative curvature, part of this reduction in the FoM ratio might be prior-induced, which itself reduces the posterior for  large and positive $\Omega_K$.}
\label{fig:pca_fom_geo2} 
\end{figure}

Figure~\ref{fig:pca_fom_geo} compares the PCA-based Figure of Merit between the analysis with the JLA compilation and the WFIRST simulated data. Both cases include BAO, geometric CMB, and local $H_0$ measurements. Current data can place a good measurement on approximately the first five principal components. The FoM for these modes is $\text{FoM}_{n=5}^{\text{PC}}\text{(SN=JLA)} \approx 10^4$, while $\text{FoM}_{n=2}^{\text{PC}}\text{(SN=JLA)} \approx 10^2$ and therefore two-parameter descriptions of $w(z)$, such as the commonly adopted $w_0-w_a$, does not exhaust the information that can be extracted from current data. Future WFIRST data will be able to better constrain the first five principal components significantly ($\text{FoM}_{n=5}^{\text{PC}}\text{(SN=WF)} \approx 10^6$), and it will also probe twice the number of principal components in comparison to JLA type IA supernovae. Indeed, WFIRST will have a wider redshift range sensitive to modes that affect $w(z)$ only at high-redshift.  The asymptotic FoM ratio between these two datasets is of order $10^4$, while this ratio is on the order of $10$ for the $w_0-w_a$ parametrization. 

In the $w_0-w_a$ functional form, the high and low-redshift are entangled, and therefore many of the modes that only WFIRST can measure well are not allowed by prior or are severely restricted by the low-redshift supernovae data. Thus, simple two-parameter functional forms of $w(z)$ offer an incomplete picture of the modes that can be measured with current and future data. Including extra parameters in popular parametrizations, however, does not guarantee that the signal-to-noise ratio of higher order PCAs will be greater than unity and, therefore, the approximate upper limit in $\text{FoM}_{n}$ may be quite difficult to achieve. This limitation indeed seems to hold, as it has been shown that the signal-to-noise ratio in $e_{n \gg 3}(z)$ PCAs, constructed to be representative of experiments that resemble WFIRST, is small in commonly adopted quintessence models~\cite{dePutter:2008bh,Barnard:2008mn,Linder:2005ne}. Without selection criteria derived from a more fundamental description of dark energy that indicates that these are the only well-motivated theories, our results merely indicate that, in principle, it is possible to construct smooth dark energy models that are elaborate enough so that improvements in $\text{FoM}^{\text{PC}}_{n}$ with $n \gg 3$ represent a gain of information provided by next generation of experiments\footnote{The PCA merely shows what are the modes that can be constrained by the data (and they show how well they can be constrained). Interpretation of the subspace spanned by these PCs depends on theoretical analysis that is out of the scope of statistical tools that rely only on the data}.

In quintessence scenarios, priors in the allowed curvature range make a significant impact in the Figure of Merit (as shown in Figure~\ref{fig:pca_fom_geo2}). Indeed, Figure~\ref{fig:curvature_geo_current} shows that curvature cannot be constrained at the few percent level with current data if we assume quintessence, and the uniform prior $|\Omega_K | < 0.1$ decreases the FoM by approximately an order of magnitude in comparison to the flat $\Omega_K = 0$ case. In general, in smooth dark energy models that respect the boundary $-2.5 < w(z) < 1$, the curvature posterior disfavors large positive values. However, the chosen uniform priors in the amplitude of the principal components is not mapped into flat spatial curvature posteriors, and in fact the priors provide more weight to models that do cross the phantom barrier (see Figure 12 of~\cite{Mortonson:2008qy}). 

The inability of current geometric data to constrain spatial curvature at the percent level when $\Omega_K$ is marginalized over quintessence and general dark energy models provides an excellent opportunity for observables that directly measure the growth of structure to make a significant impact on the Figure of Merit. Growth information could also mitigate, in the context of $\Lambda$CDM, the impact of the discrepancy between the CMB and local $H_0$ measurements, which shifts the spatial curvature posterior towards positive values. The combination of the full CMB temperature and polarization power spectra, CMB lensing reconstruction, redshift space distortions, and weak lensing measurements should indeed constrain curvature tightly given that we see a strong correlation between $\Omega_K$ and the predicted $\sigma_8 \Omega_m^{1/2}$ in all our {\it Geo} chains with current JLA supernovae (see Figure~\ref{fig:curvature_geo_current}). Indeed, the combination $\sigma_8 \Omega_m^{1/2}$ corresponds to the direction in parameter space that is best measured by weak lensing. 

With the simulated WFIRST supernovae data, $\Omega_K$ is constrained at the percent level in both quintessence and smooth dark energy scenarios (see Figure~\ref{fig:curvature_geo_future}), and this provides an interesting challenge for the future WFIRST weak lensing survey. Current weak lensing surveys, including the recently published Year One Dark Energy Survey (DES) measurements, are discrepant with CMB inferences on $\sigma_8 \Omega_m^{1/2}$ at the two sigma level. Also note that our results do not imply that every single quintessence and smooth dark energy model will have $\Omega_K$ uncertainties larger than a percent. What we show is what happens when we are agnostic concerning the feasibility of arbitrary complicated smooth dark energy scenarios.

Finally, the model-independent $\text{FoM}_n^{\text{PC}}$ can be converted into model-based Figure of Merit evaluations with the use of a fast approximate likelihood that dispenses the use of expensive additional MCMC calculations as well as the use of sophisticated numerical packages such as CosmoMC. Given the discrete set of parameter values $\boldsymbol{\alpha}_i=\{\alpha_1,...,\alpha_{20}\} $ and multiplicities $w_i$ provided by our MCMC chains, we define a kernel density estimation likelihood of the form~\cite{Heinrich:2016ojb}
\begin{align}
\mathcal{L}_{\text{PC}}(\text{data}|\boldsymbol{\alpha}) = \sum_{i=1}^{N} w_i K_f (\boldsymbol{\alpha}-\boldsymbol{\alpha}_i).
\end{align}
Here $N$ is the number of elements in the chain, $K_f$ is a smoothing kernel that we assume to be a multivariate Gaussian with zero mean and covariance $f \matr{C}_{n=20}$ (f is a smoothing factor), and $\boldsymbol{\alpha}$ is the set of values generated by the model to be constrained. Such technique has been applied with remarkable success in the context of model-independent studies on the epoch of reionization~\cite{Heinrich:2016ojb,Miranda:2016trf}. The posterior for any physically motivated parametrization with M parameters $\boldsymbol{\beta}=\{\beta_1,...,\beta_{M}\}$ is
\begin{align}
P (\boldsymbol{\beta} |\text{data}) \propto  \mathcal{L}_{\text{PC}}(\text{data}|\boldsymbol{\alpha}(\boldsymbol{\beta})) P(\boldsymbol{\beta}) \,.
\end{align}
We then can define model-based FoM as $\text{FoM}_M^{\text{Model}} = \big(\text{det} \,\, \matr{C}(\beta_1,...,\beta_M)\big)^{-1/2}$. Such posterior also allow the signal-to-noise ratio of each principal component to be fully sampled for arbitrary models. We intend to fully explore this technique in a future work to be accomplished in collaboration with the WFIRST supernova science investigation teams.


\section{Improving Spatial Curvature constraints with growth information}
\label{sec:growth_probes}

In this section, we will use the PCA basis of $w(z)$ to quantify the effects of marginalizing the spatial curvature posterior over different classes of dark energy models. We will restrict our analysis to current data, including information from the growth of structure. In a follow-up study, we will investigate the correlations between dark energy parameters and spatial curvature for the future WFIRST mission, including the WFIRST weak lensing survey. We will also quantify the correlations between the sum of neutrino masses constraints and dark energy scenarios. 

In the context of the $\Lambda$CDM model, the combination of the full CMB temperature and polarization spectra from the Planck satellite with BAO measurements constrains the spatial curvature to the sub-percent level $|\Omega_K| < 0.005$ \cite{Ade:2015xua}. Figure~\ref{fig:curvature_geo_current} shows that even datasets that only measure the background expansion of the universe can, all together, probe the curvature to within a percent. However, the discrepancy between the CMB and local $H_0$ measurements shifts the central value of the spatial curvature posterior towards positive values. For quintessence models, Figure~\ref{fig:curvature_geo_current} shows that geometric data cannot constrain spatial curvature even at the ten percent level. 
The situation changes slightly for general smooth dark energy models for which  $-2.5 < w(z) < 1$. However, this shift towards negative spatial curvature is in part due to our choice of priors, given that uniform priors in the PCA amplitudes $\alpha_i$ do not translate into a flat posterior in the spatial curvature \cite{Mortonson:2008qy}.  

\begin{figure}[t]
\includegraphics[scale=0.44]{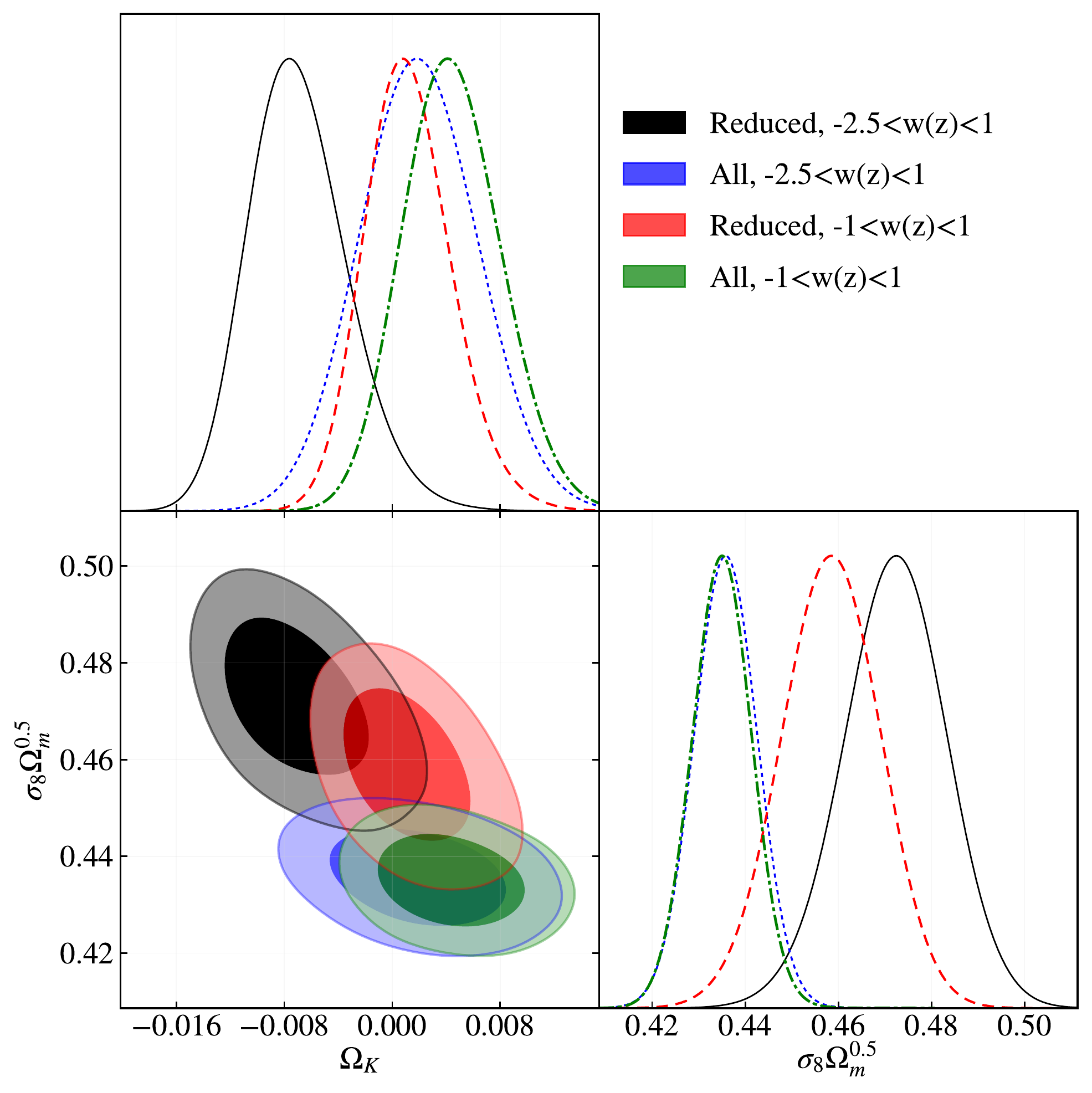}
\caption{Posterior of curvature and its correlation with $\sigma_8\Omega_m^{1/2}$. It is clear that the tension between CMB measurements by the Planck satellite and low redshift structure probes shifts $\Omega_K$ at the two sigma level. In particular, no claims that curvature is measured at the sub-percent level, when marginalized over smooth dark energy models, can be made without solving this tension.}
\label{fig:curvature_s8}
\end{figure}

The remaining freedom in the spatial curvature posterior can be significantly reduced by constraining the product $\sigma_8 \Omega_m^{1/2}$, which is a combination of parameters that weak lensing measures best. CMB temperature and polarization power spectra can also constrain the curvature up to percent level due to the gravitational lensing effect that smooths the acoustic peaks. Indeed, Figure~\ref{fig:curvature_s8} shows that the posterior width of $\Omega_K$ is reduced by more than a factor of $10$ in the context of quintessence models when the Gaussian CMB likelihood is replaced by the full CMB temperature and polarization spectra. Even for arbitrary smooth dark energy models, with $-2.5 < w(z) < 1$, the spatial curvature can be constrained better than $1.5\%$ with the full CMB power spectrum. 

In both quintessence and smooth dark energy paradigms, the effect of marginalizing the dark energy principal components in addition to spatial curvature is to reduce the FoM by a factor of few. More specifically, the FoM is reduced by $\approx 10$ in quintessence scenarios, primarily because of the non-negligible posterior probability for large and positive $\Omega_K$, and by $\approx 5$ in smooth dark energy scenarios, as shown in Figure~\ref{fig:pca_fom_geo2}.  This level of improvement is significantly higher than in $\Lambda$CDM. Indeed, the spatial curvature is constrained by both geometry and growth information in the standard model, given that there is not enough freedom in the dark energy sector to compensate changes in the background expansion induced by large values of $\Omega_K$ to maintain the comoving distance to the surface of the last scattering unchanged.

The addition of low-redshift probes that measure growth, such as weak-lensing and redshift space distortion, shifts the curvature posterior by an amount comparable to the $95\%$ confidence regions, in comparison to the {\it Reduced} MCMC chains (see Figure~\ref{fig:curvature_s8}). Indeed, the combination $\sigma_8\Omega_m^{1/2}$ is well constrained by weak lensing, and there is a two sigma tension between weak lensing and the CMB, which reflects into doubling the uncertainties in constraining spatial curvature marginalized over smooth dark energy scenarios. 

\section{Discussion}
\label{sec:discussion}

In this paper, we provide a comprehensive investigation on how current data that probe the background expansion constrain the theoretical predictions of three broad classes of dark energy: $\Lambda$CDM, quintessence, and smooth dark energy models that respect the prior $-2.5 < w(z) < 1$. These three paradigms share the property that dark energy influences the growth of structure by modifying the background expansion. 

Within this framework, we show that the current background expansion predicts the linear growth of structure at the percent level. For general smooth models, such predictions are at the $10$ percent level when marginalizing over the informative prior $-0.01 < \Omega_k < 0.01$. Flat models always predict growth at the few percent level, which provides an exciting opportunity for current and future surveys to falsify the flat, smooth dark energy scenario with weak lensing and redshift-space distortion measurements. 

WFIRST supernovae data will be able to improve growth predictions in curved models significantly. In particular, the two-sigma posterior for $G(z)$ is at the $8\%$ level even when marginalizing over the non-informative prior $-0.1 < \Omega_k < 0.1$. Our analysis is conservative because it neglects upcoming BAO improvements from the future DESI survey as well as advancements in measuring the local $H_0$. 

In the near future, the degeneracy between spatial curvature and $w(z)$ could be mitigated with measurements of the combination $\sigma_8 \Omega_m^{1/2}$. Indeed, this is the direction in parameter space that weak lensing measurements restrict the most. In fact, the inclusion of CFHTLens and DR12 RSD measurements reduces the Figure of Merit of the PCA amplitudes by order unity. We also point out that inconsistencies between low-redshift measurements and CMB predictions for $\sigma_8 \Omega_m^{1/2}$ translate into uncertainties in constraining $\Omega_k$ marginalized over the PCA amplitudes.

Finally, we evaluate a PCA-based Figure of Merit, which reveals that a two-parameter description of $w(z)$ may not provide the complete picture of advancements in constraining power between WFIRST and JLA supernovae surveys. In particular, specific $w(z)$ functional forms may bias the determination of the optimal redshift range for the WFIRST supernovae survey. While a shallow survey can provide better statistics, a more extensive range may probe a broader range of models.

\acknowledgments

We thank R. Hounsell, D. Scolnic, R. J. Foley and R. Kessler for helpful discussions and for providing us state-of-the-art WFIRST simulated supernova. We also thank R. Kessler and W. Hu for providing us extensive computational resources granted by the University of Chicago Research Computing Center. We thank Eric Linder for valuable discussions about the interpretation of the PCA-based Figure-of-Merit. VM  thanks the hospitality of the Sitka Sound Science Center during the intermediate stages of this work. VM was supported in part by the Charles E.~Kaufman Foundation, a supporting organization of the Pittsburgh Foundation.

\appendix
\label{sec:appendix_GSR}

\bibliography{PCA}

\end{document}